\documentclass[fleqn,usenatbib]{mnras}

\usepackage{newtxtext,newtxmath}

\usepackage[T1]{fontenc}

\DeclareRobustCommand{\VAN}[3]{#2}
\let\VANthebibliography\thebibliography
\def\thebibliography{\DeclareRobustCommand{\VAN}[3]{##3}\VANthebibliography}

\usepackage{graphicx}	
\usepackage{amsmath}	
\usepackage{CJK}
\usepackage{bm}
\usepackage{orcidlink}
\usepackage{multirow}
\usepackage{booktabs}
\usepackage{hyperref}

\title[SPS method: SEW]{SEW: A full-spectrum linear fitting with stellar population synthesis method Based on ``Equivalent Widths spectrum''}

\author[J.F. Lu et al.]{
Jiafeng Lu(卢家风)\orcidlink{0000-0002-8817-4587}$^{1}$\thanks{E-mail: jefferslu@live.com},~
Xi Kang (康熙)\orcidlink{0000-0002-5458-4254}$^{1,2,3}$\thanks{E-mail: kangxi@zju.edu.cn},~
Shiyin Shen (沈世银)\orcidlink{0000-0002-3073-5871}$^{4,5}$~
\\
$^{1}$Institute for Astronomy, School of Physics, Zhejiang University, Hangzhou 310027, China\\
$^{2}$Center for Cosmology and Computational Astrophysics, Zhejiang University, Hangzhou 310027, China\\
$^{3}$Purple Mountain Observatory, 10 Yuan Hua Road, Nanjing 210034, China\\
$^{4}$Shanghai Astronomical Observatory, Chinese Academy of Sciences, 80 Nandan Road, Shanghai 200030, China\\
$^{5}$Key Lab for Astrophysics, Shanghai, 200034, People's Republic of China\\
}

\date{Accepted XXX. Received YYY; in original form ZZZ}

\pubyear{\the\year{}}

\begin{document}
\begin{CJK*}{UTF8}{gbsn}
\label{firstpage}
\pagerange{\pageref{firstpage}--\pageref{lastpage}}
\maketitle

\begin{abstract}
We present a full-spectrum linear fitting method, SEW, for stellar population synthesis based on equivalent widths (EWs) to extract galaxy properties from observed spectra. This approach eliminates the need for prior assumptions about dust attenuation curves, which are instead derived as outputs of the fitting process. By leveraging the invariance of EWs and employing the Discrete Penalised Least Squares (DPLS) method to extract EWs, we address the nonlinear aspects of the fitting process by linearising the matrix equations. This enables accurate recovery of key parameters, stellar age, metallicity and dust attenuation, even under systematic calibration biases and varying attenuation conditions. Rigorous testing with mock spectra across signal-to-noise ratios (S/N = 5-30) and calibration biases demonstrates the robustness of method. The derived attenuation curves align closely with input models, and stellar population parameters are recovered with minimal bias. To facilitate adoption, we implement this method as a Python extension package for \texttt{pPXF} (\texttt{pPXF-SEW}). Our work addresses critical degeneracies in traditional spectral fitting and enhances the reliability of extragalactic studies.
\end{abstract}




\section{Introduction} \label{sec:intro}

Galaxy spectra analysis is a cornerstone in astrophysics, providing insights into the physical and evolutionary characteristics of galaxies. However, it faces several challenges that complicate the accurate extraction of key parameters such as star formation history, metallicity, and dust content. Dust attenuation is a significant issue, as dust within and around galaxies absorbs and scatters light, causing wavelength-dependent dimming and reddening, which complicates the determination of intrinsic stellar properties; the selection of an appropriate dust attenuation curve is crucial but often uncertain, with different models producing significantly varied results \citep[e.g.,][]{Fitzpatrick1999, Calzetti1994, CF00}. Systematic calibration biases from flux calibration inaccuracies lead to incorrect absolute brightness measurements, and fitting templates, such as simple stellar population (SSP) models, may not perfectly represent real stellar populations, introducing errors \citep{Conroy2009}. Degeneracy among parameters, including age, metallicity, and dust content, is problematic, as their similar effects on the spectrum make them difficult to distinguish, with young, metal-poor, dusty galaxies potentially having spectra resembling those of old, metal-rich, less dusty ones, leading to large uncertainties, particularly with traditional full-spectrum fitting methods \citep{Worthey1999, Conroy2013}.

In Galaxy spectra analysis, two primary stellar population synthesis (SPS) methods are commonly employed: the full-spectrum fitting and the absorption line characterisation method.

In full-spectrum fitting, observed spectra are modelled through linear combinations of SSPs, each characterised by distinct age and metallicity parameters. These SSP templates incorporate integration over an initial mass function (IMF) and stellar evolutionary tracks. The fitting process adjusts the weights of SSPs, along with other parameters, to minimise residuals between the model and data: 
\begin{subequations}
\begin{align}
    F(\lambda) & = F_{\text{intr}}(\lambda) * P(\lambda), \label{eq:f_fintr} \\
    F_{\text{intr}}(\lambda) & = \sum_j x_{j} f_{\text{temp},j}(\lambda), \label{eq:csp}\\
    P(\lambda) & = 10^{-0.4A(\lambda)} * q(\lambda). \label{eq:att+q}
\end{align}
\end{subequations}
Here, $ F_{\text{intr}}(\lambda)$  represents the intrinsic, unattenuated galactic spectrum comprising SSP and emission-line templates $ f_{\text{temp}}^{(j)}(\lambda) $ with corresponding weights $ x_j $. The observed spectrum $ F(\lambda) $ emerges through modification by the global variations $ P(\lambda) $, which accounts for both intrinsic (self-) and foreground Galactic extinction $A(\lambda)$, combined with instrumental calibration deviations $q(\lambda)$. This inverse modelling approach enables the recovery of intrinsic galactic properties from observational data through spectral decomposition. Several established spectral synthesis codes implement this methodology, including \texttt{pPXF} \citep{Cappellari2004, Cappellari2017}, STARLIGHT \citep{CidFernandes2005}, STECKMAP \citep{Ocvirk2006a, Ocvirk2006b}, and VESPA \citep{Tojeiro2007}. However, the accuracy of the fitting results is significantly influenced by several interrelated factors, including the choice of the theoretical dust attenuation curve, the precision of spectral calibrations, and the inherent degeneracy among age, metallicity, and dust properties. These elements collectively affect the shape of the continuum spectrum. The introduction of additional parameters to account for systematic biases and correct for dust attenuation further exacerbates this complexity.

On the other hand, absorption line characterisation method, such as the Lick/IDS indices \citep{Burstein1984, Worthey1994}, focuses on the equivalent width of specific absorption features in the galaxy spectrum. These features, including Mg b, Fe, and H$\beta$ lines, are sensitive to the ages and metallicities of stellar populations. By comparing the measured line indices to those predicted by stellar population models, one can derive the average age, metallicity, and other properties of the stars in the galaxy \citep{Tripicco1995}. For instance, \citet{Trager2000} applied these indices to study the stellar populations of elliptical galaxies, and \citet{Gallazzi2005} used them for analyzing the stellar populations in Sloan Digital Sky Survey (SDSS) data. Particularly, the full-index fitting (FIF), which compares the flux strength within a specific absorption feature pixel by pixel rather than averaging them, effectively breaks age-metallicity-IMF degeneracies \citep{Martin2015,Martin2019, Ditrani2025}. These approaches of equivalent width indices and FIF are less susceptible to dust attenuation and flux calibration biases, as their normalization and localization inherently exclude the continuum variations (as discussed in Section \ref{sec:inv of ew}). However, these methods remain limited by signal-to-noise and resolution requirements, typically only utilizing a few dozen strong absorption lines among tens indices defined across the UV-to-IR. While these methods have driven major advances in understanding galaxy stellar populations, their limited spectral sampling may hinder comprehensive property assessments. Consequently, such methods may have limitations in fully determining intrinsic spectral variations and extinction details within galaxies.

By extending the principles behind the absorption line characterisation method, it is conceivable that each pixel (whether or not absorption line features exhibit) in the spectrum can be associated with an ``equivalent absorption line''. This approach significantly broadens the applicability of absorption line characterisation, from a mere handful of lines to the entirety of the spectrum. This method integrates the benefits of full-spectrum fitting and absorption line characterisation, not only effectively overcoming the degenerate posed by dust attenuation and flux systematic bias. but also leveraging the rich information embedded in the entire spectrum. \citet{Wilkinson2015,Wilkinson2017} first developed a code of full spectral fitting (Firefly) based on the premise that dust attenuation and systematic bias primarily influence the large-scale structure of the continuum spectrum, while sparing the small-scale structure of absorption lines. they employed a Fourier-based high-pass filter method (HPF), effectively stripping the observed and model spectra of large-scale features. However, it is crucial to recognize that the HPF approach alone cannot completely eliminate the residual signal from dust attenuation and systematic bias within the small-scale features. To isolate small-scale spectral features from large-scale variations, \cite{Liniu2020} utilise a moving box average. By analyzing the ratio between these features, they are able to constrain the underlying stellar population. Since dust attenuation impacts both scales uniformly at a given wavelength, the ratio remains unaffected by dust attenuation and systematic bias. However, this method is nonlinear, thereby constraining the parameter space and reducing the operational speed when addressing stellar populations.

In this work, we present a full-spectrum linear fitting method based on equivalent widths. This method do not need a prior attenuation curve assumption and calibration bias assumption, while facilitating rapid fitting utilising full-spectrum information due to its linear nature. In order to enhance usability, we have developed this method as a Python package. The package has been released in PYPI \footnote{\url{https://pypi.org/project/ppxf-sew/}}.

The outline of this paper is as follows. In Section \ref{sec:method}, we introduce the full-spectrum linear fitting method based on equivalent widths. Then, we we test the performance of our method with mock spectra and statistical tests in Section \ref{sec:mock}. Finally, we make some discusses in Section \ref{sec:discussion} and summary in Section \ref{sec:summary}.

\section{METHOD} \label{sec:method}

In this section, we present the central concept and methodology utilised in the full-spectrum linear fitting with stellar population synthesis method based on ``Equivalent Widths spectrum'' (SEW). Specifically, we discuss the principle that equivalent width can be regarded as invariant, describe the method for deriving the full-spectrum equivalent width, outline the method for linearising the equivalent width equation, detail the approach for determining the global variation and attenuation, and explain the procedure employed for kinematic analysis and other fitting processes.

\subsection{Invariance of Equivalent Width}
\label{sec:inv of ew}

The equivalent width ($EW$) is the width of the adjacent continuum $F'(\lambda)$ that has the same area as is taken up by the absorption or emission line in original spectrum $F(\lambda)$. Mathematically we have:
\begin{equation}
    EW=\int \frac{F(\lambda)-F'(\lambda)}{F'(\lambda)} d\lambda,
\end{equation}
where $F'$ is the background smoothed continuum spectrum\footnote{The superscript ' in the following text indicates the smoothing continuum of the corresponding spectrum.}. Similarly, we can define the full-spectrum ``equivalent width spectrum'' as:
\begin{equation}
    EW(\lambda)=\frac{F(\lambda)}{F'(\lambda)}.
\end{equation}
It is worth noting that we are unable to directly observe the "true" continuum, $F'(\lambda)$. Typically, we resort to defining a "pseudo-continuum" by utilising neighbouring regions or smoothed spectra. 

The continuum $F'(\lambda)$ depends on the original spectrum $F(\lambda)$. A well-defined continuum effectively mirrors any global variations $P(\lambda)$ in the shape of the original spectrum, encompassing factors like dust attenuation and calibration bias:
\begin{equation}
    F'(\lambda) = F'_{intr}(\lambda)*P(\lambda),
\label{eq:smo_p}
\end{equation}
where the $F'_{intr}(\lambda)$ is the continuum of intrinsic original spectrum $F_{intr}(\lambda)$ that this equation corresponds to Equation \ref{eq:f_fintr}.

It can be seen that, a well-defined Equivalent Width is invariant in global variations $P(\lambda)$:
\begin{equation}
    EW(\lambda)=\frac{F(\lambda)}{F'(\lambda)}=\frac{F_{intr}(\lambda)*P(\lambda)}{F'_{intr}(\lambda)*P(\lambda)}=\frac{F_{intr}(\lambda)}{F'_{intr}(\lambda)}.
\label{eq:ew}
\end{equation}

The invariance of the equivalent width is specifically applicable to the equivalent width spectrum, rather than the continuum-subtracted residual spectrum. 
\begin{equation}
    RES(\lambda)=F(\lambda)-F'(\lambda)=(F_{intr}(\lambda)-F'_{intr}(\lambda))*P(\lambda).
\end{equation}
The latter retains the attenuation and calibration bias. Consequently, invariance cannot be assured for the continuum-subtracted residual spectrum, which makes it less suitable for direct application in stellar population synthesis.

The invariance of the equivalent width is established on the premise that global variability remains consistent across all stellar populations. Instrumental effects lead to calibration bias, a global phenomenon that uniformly affects all stellar populations. However, different stellar populations may encounter varying degrees of attenuation, which introduces additional uncertainties. A more detailed discussion on this topic will be presented in Section \ref{sec:inconsistency}.

\subsection{Discrete Penalised Least Squares}
\label{sec:dpls}
In our method, in order to obtain the equivalent width spectrum, we first need to extract the continuous spectrum, i.e., smooth the spectrum. We apply the discrete penalised least squares method to smooth \citep{paul2003}. The general idea behind this smoothing method is to make the smoothed spectrum $F'$ match the original spectrum $F$ as well as it can while also penalizing the roughness of the smoothed data. 

Specifically, this method involves balancing two opposing goals: (1) fidelity to the original data and (2) smoothness of $F'$. The smoother $F'$ is, the more it will deviate from $F$. Smoothness of $F'$ can be expressed with differences $\Delta F'_i$ or Higher-order differences $\Delta^{n} F'_i$. Sometimes, the spectrum are not sampled at equal intervals. For example, some observations are linear in logarithmic wavelength space, which means that the pixel spacing is nonuniform sampling. In the case, the $\Delta^{n} F'_i$ is divided differences, for example:
\begin{equation}
\begin{aligned}
    \Delta F'_{i}&=\frac{F_{i}-F_{i-1}}{x_{i}-x_{i-1}} \\
    \Delta^2 F'_{i}&= \frac{\Delta(\Delta F_{i})}{x_{i}-x_{i-2}}=\frac{\frac{F'_{i}-F'_{i-1}}{x_{i}-x_{i-1}}-\frac{F'_{i-1}-F'_{i-2}}{x_{i-1}-x_{i-2}}}{x_{i}-x_{i-2}} \\
    \Delta^n F'_{i}&=\frac{\Delta(\Delta^{n-1} F'_{i})}{x_{i}-x_{i-n}}. 
\end{aligned}
\end{equation}
In this work, we set $n$ to 2, when the divided differences performs best as a penalty term \citep{paul2003}, which is the discrete form of the second-order derivative or curvature. For spectra, the x of divided differences is $\log \lambda$ here in order to ensure that smoothing and spectral line broadening keep uniform at different wavelengths. A simple and effective measure of the smoothness of $F'$ can be gives by the sum of squares of the differences: $R = \sum_{i}(\Delta^n F_{i})^{2}$. The lack of fit to the data can be measured as the usual sum of squares of differences: $S = \sum_{i} (F_{i}- F'_{i})^{2}$. The key lies in finding a balanced combination of these two goals, achieved by minimizing the sum $Q = S + \sigma R$, where $\sigma$ is a user-defined parameter. The essence of penalised least squares is to determine the series $F'$ that minimises Q. As the penalty scale factor $\sigma$ increases, the influence of roughness R on the overall objective Q becomes more significant, resulting in a smoother $F'$ at the cost of reduced fidelity to the original data $F$. In practice, the choice of the $\sigma$ for the smoothing of galaxy spectra is discussed in detail in Section \ref{sec:sigma}.

When extracting a continuous spectrum, some pixels are quite often unavailable, such as emission line pixels and masked pixels. The smoother can easily be modified to handle this situation. A vector $\bm{w}$ of weights is introduced, with $w_i = 0$ for unavailable pixels and $w_i = 1$ otherwise. This ensures us handle missing values, even in large range.

Furthermore, this method exhibits a remarkable ability to adapt seamlessly to spectral boundaries \citep{paul2003}, significantly enhancing spectrum utilization. In particular, when the resting wavelength boundary of the spectrum falls within the range of 3500-4200  \r{A}, this method leverages the absorption features of Balmer and Ca KH close to the boundary, thereby bolstering the precision of stellar population measurements. Conversely, the Fourier and moving box average methods encounter significant challenges at the boundaries \citep{paul2003}, thereby diminishing the overall spectral usability.

The resulting general function that is minimised to determine the smooth data is then:
\begin{equation}
    Q= \sum_{i}^{N} w_i (F_{i} - F'_{i})^{2} + \sigma \sum_{i}^{N - n} (\Delta^{n} F'_{i})^{2}
\end{equation}
where $F$ is the original data, $F'$ is the smoothed data, $\sigma$ is the penalty scale factor, $w_i$ is the weighting.

To avoid a lot of tedious algebra, the matrix and vector forms are follows:
\begin{equation}
    Q=(\bm{F}-\bm{F'})^{\top}\bm{W}(\bm{F}-\bm{F'})+\sigma|\bm{D_n F'}|^2, \,
\end{equation}
where the superscript $\top$ denotes the transpose of the matrix, $\bm{W}$ is a diagonal matrix with $\bm{w}$ on its diagonal, and $ \bm{D_n}$ is a band matrix of order n such that $ \bm{D_n F'} = \Delta^n \bm{F'}$. For example, for an uniform sampling data of length 5, the  $\bm{D_2}$ is:
\begin{equation}
\begin{aligned}
    \bm{D_2} &= \begin{pmatrix}
                1 & -2 & 1 & 0 & 0 \\
                0 & 1 & -2 & 1 & 0 \\
                0 & 0 & 1 & -2 & 1 \\
                \end{pmatrix}                         
\end{aligned}
\end{equation}

Using results from matrix calculus, we find for the vector of partial derivatives:
\begin{equation}
    \frac{\partial Q}{\partial \bm{F'}}= -2 \bm{W} (\bm{F}-\bm{F'})+ 2\sigma\bm{D_n}^{\top} \bm{D_n F'}, \,
\end{equation}
and equating this to 0, we get the linear system of equations:
\begin{equation}
    (\bm{W} + \sigma \bm{D_n}^{\top} \bm{D_n}) \bm{F'} = \bm{W F}. \,
\label{eq:wf}
\end{equation}

As mentioned in Section \ref{sec:inv of ew}, a well-defined continuum should reflects any global variations in the shape of the original spectrum. In other words, The intrinsic continuum $F'_{intr}$ corresponding to the smoothed continuum $F'$ derived from Equation \ref{eq:wf}, adheres to the same Equation \ref{eq:wf} when related to the intrinsic continuum $F_{intr}$.

For simplify, let us designate $\bm{W} + \sigma \bm{D_n}^{\top} \bm{D_n}$ as matrix $\bm{A}$. According to the definitions of $\bm{D_n}$ and $\bm{W}$, $\bm{A}$ is a banded matrix\footnote{A matrix in which all non-zero elements are concentrated in a band centered on the main diagonal.}.  By combining Equations \ref{eq:f_fintr}, \ref{eq:smo_p}, and \ref{eq:wf}, we arrive at:
\begin{equation}
    \bm{APF'_{intr}}=\bm{WF}=\bm{WPF_{intr}}  
\end{equation}
since both $\bm{W}$ and $\bm{P}$ are diagonal matrices, we can rearrange the equation to:
\begin{equation}
    \bm{P^{-1}APF'_{intr}}=\bm{WF_{intr}} 
\end{equation}
where $\bm{P^{-1}}$ is the inverse matrix of the diagonal matrix $\bm{P}$. Because $\bm{P}$ is diagonal, elements $a'_{i,j}$ of $\bm{P^{-1}AP}$ can be written as:
\begin{equation}
    a'_{i,j}=\frac{1}{p_{i,i}}a_{i,j}p_{j,j}.
\end{equation}
Notably, the matrix $\bm{A}$ is a band matrix, meaning that its elements are concentrated near the diagonal. Therefore, $p_{i,i}$ and $p_{j,j}$ are elements located in close proximity. Since $\bm{P}$ describes global variations, the values of $p_{i,i}$ and $p_{j,j}$ only exhibit minor variations, i.e., $a'_{i,j} \approx a_{i,j}$. This implies that $\bm{P^{-1}AP} \approx \bm{A}$.

\subsection{Linearization of Equation}
\label{sec:linear}
In our method, we fit the equivalent width of observed spectrum (Equation \ref{eq:ew}) with that of the model spectrum:
\begin{equation}
    \bm{EW}=\frac{\bm{F_{temp}} x}{\bm{F'_{temp}} x},
\end{equation}
where $F_{temp}$ and $F'_{temp}$ are the template matrix and the smoothed template matrix, respectively. $\bm{EW}$ is observed equivalent width. $x$ is the component weights to fit, consisting of two parts, the stellar population component $x_{ssp}$ and emission line component $x_{em}$. It is evident that this matrix equation is nonlinear, preventing us from rapidly solving for $x$ using a linear matrix. In \cite{Liniu2020}, the authors utilise a nonlinear least-squares fitting code, MPFIT, and generate a small set of model templates by employing Principal Component Analysis (PCA). Their approach requires significant computational resources and is limited in terms of the number of parameters it can handle. In this section, we aim to linearise this system.

First, the equivalent width spectrum can be regarded as a composition of the emission line equivalent width spectrum and a composition of stellar equivalent width spectrum, both of which are equivalized by the stellar continuum:
\begin{equation}
\begin{aligned}
    &\bm{EW}=\bm{EW_{csp}}+\bm{EW_{em}}=\frac{\bm{F_{csp}}+\bm{F_{em}}}{\bm{F'}} \\
    &where\\
    & \bm{EW_{csp}} =\frac{\bm{F_{csp}}}{\bm{F'}}=\frac{\bm{F_{ssp}}x_{ssp}}{\bm{F'_{ssp}}x_{ssp}},\\
    & \bm{EW_{em}} = \frac{\bm{F_{em}}x_{em}}{\bm{F'}}.   
\end{aligned}
\label{eq:ew_matrix}
\end{equation}

We note that the stellar component can be rewritten by the following form:
\begin{equation}
\begin{aligned}
        &\bm{EW_{csp}}=\frac{\bm{F_{ssp}} x_{ssp}}{(\bm{F'_{ssp}}-1) x_{ssp}+sum(x_{ssp})},\\
        &then \\
        &\bm{F_{ssp}} x_{ssp}=\bm{EW_{ssp}}(\bm{F'_{ssp}}-1) x_{ssp}+\bm{EW_{csp}} sum(x_{ssp}), \\
        &then \\
        &(\bm{F_{ssp}} - \bm{EW_{csp}} (\bm{F'_{ssp}}-1))x'_{ssp} = \bm{EW_{csp}},
\end{aligned}
\end{equation}
where $x'_{ssp}$ is the normalized weight vector that the sum of $x'_{ssp}$ equals to 1. This form of the equation ensures it have a non-zero least square solution. 

Finally, the total equivalent width spectrum is written as:
\begin{equation}
\begin{split}
    \bm{EW} = \begin{bmatrix}
        \bm{F_{ssp}} - \bm{EW_{csp}} (\bm{F'_{ssp}}-1) \\
        \frac{\bm{F_{em}}}{\bm{F'}}
    \end{bmatrix} x_{tot},
\end{split}
\label{eq:block}
\end{equation}
where the square brackets denote a block matrix consisting of two internal sub matrices. In this equation, the $\bm{EW}_{csp}$ is unknown. One possible way is to guess a $x'_{ssp}$ by replacing $\bm{EW}_{csp}$ with $\bm{EW}$, since $\bm{EW} \sim \bm{EW}_{csp}$. Then use the guessed $x'_{ssp}$ to obtain $\bm{EW}_{csp}$ through equation \ref{eq:ew_matrix}. Eventually, a final $x'_{ssp}$ is obtained through several rounds of iteration. For most cases in our tests, no more than five rounds of iterations result in a constant $x'_{ssp}$.

In addition, it requires that the stellar population weights $x \ge 0$. This can be solved by methods such as quadratic programming, or nonnegative least squares. Similar methods are widely used in many stellar population synthesis code such as pPXF, and will not be repeated here.

\subsection{Global variation and attenuation}
\label{sec:var&att}

Once the stellar population weights, $x'_{ssp}$, are obtained, the best-fitting spectrum, without global bias and dust attenuation, can be reconstructed as follows: 
\begin{equation}
F_{fit}(\lambda) = \sum_j x'_{ssp} f_{temp}(\lambda).
\end{equation}
Combining this with Equation \ref{eq:f_fintr}, the global correction function can be evaluated as: 
\begin{equation}
P(\lambda) = \frac{F(\lambda)}{f_0 \cdot F_{fit}(\lambda)}.
\end{equation}
Here, $f_0$ is a normalization factor to adjust the absolute value of the spectrum, which is unknown. In practice, we relate the global correction function to the V-band ($\lambda_V = 5500$ \r{A}) to eliminate $f_0$: 
\begin{equation}
    \frac{P(\lambda)}{P_V} =  \frac{F(\lambda)/F_{fit}(\lambda)}{F(\lambda_V)/F_{fit}(\lambda_V)}.
\end{equation}
The global correction function $P(\lambda)$, as described in Equation \ref{eq:att+q}, consists of two components: calibration bias and attenuation. Therefore, the equation can be written following the defination of attenuation as:
\begin{equation}
    \begin{aligned}
    -2.5 \log \frac{P(\lambda)}{P_V} &=  A(\lambda) -A_V -2.5 \log \frac{q(\lambda)}{q_V} \\
    &= -2.5 (\log \frac{F(\lambda)}{F_{fit}(\lambda)} - \log \frac{F(\lambda_V)}{F_{fit}(\lambda_V)}).  
    \label{eq:alam-av}
    \end{aligned}  
\end{equation}
It is important to note that both calibration bias and attenuation influence the overall characteristics of the spectra. These effects can only be disentangled using parametric methods, such as assuming attenuation curves. In our approach, we do not attempt to separately decompose these two components; rather, we consider them collectively in our output. 

Specifically, under the condition of accurate spectral calibration where $q(\lambda) = 1$, $P(\lambda)$ directly quantifies spectral attenuation. This curve allows for a detailed investigation of attenuation curves of both global morphology and broad absorption features, such as the 2175\r{A} UV bump and the intermediate scale structure\citep{Massa2020,Zhang2024}. Furthermore, it provides a foundation for exploring the physical mechanisms underlying dust attenuation processes.

\subsection{The SEW Extension for pPXF}

In stellar population synthesis, it is crucial to account for the broadening of spectral absorption lines and the spectral shift effects caused by the intrinsic kinematics of galaxies as well as instrumental effects. Both of these effects are nonlinear. Generally, the broadening induced by instrumental effects is corrected using the line spread function (LSF) applied to the template spectrum, while the observed spectrum results from the convolution of the stellar population spectrum and the line-of-sight velocity distribution(LOSVD) function, which describes the velocities, velocity dispersions, and other kinematics of galaxies.

\texttt{pPXF}\footnote{\url{https://www-astro.physics.ox.ac.uk/~cappellari/software/\#sec:ppxf}}\footnote{\url{https://pypi.org/project/ppxf/}}\citep{Cappellari2004,Cappellari2017}, a Python-based package for stellar population synthesis, provides a comprehensive set of methods for processing line-of-sight velocity distributions. By leveraging extensibility of Python, we address the aforementioned nonlinear processes through dependencies on \texttt{pPXF} as a package. This way, we can extend its functionality through inheritance and subclassing it. Therefore, we refer to our code as \texttt{pPXF-SEW}, an extension package of \texttt{pPXF} utilising the SEW method. Specifically, our code functions as a subclass of pPXF, inheriting most of its attributes and methods while incorporating new techniques for extracting equivalent width spectra from both observed and template spectra, as detailed in Section \ref{sec:dpls}. Furthermore, we override the methods for managing linear processes in \texttt{pPXF} using the approaches outlined in Section \ref{sec:linear} and introduce the attenuation curve attribute discussed in Section \ref{sec:var&att}. 

It is noteworthy that the linearization matrix must be recomputed at each iteration of the nonlinear process. Consequently, our method demands approximately 2 to 3 times the computational time compared to the traditional approach in \texttt{pPXF}.

Since our code inherits from \texttt{pPXF}, most of its input types, output types, and parameter settings remain consistent with those in pPXF. The key difference is that \texttt{pPXF} requires a prior attenuation curve for full spectrum fitting, while our code uses the equivalent width spectrum for fitting, making the attenuation curve an output of the code rather than a prior input.

Nevertheless, our code utilises \texttt{pPXF} to handle nonlinear processes, and we recommend that anyone using the \texttt{pPXF-SEW} code cite the relevant work of \texttt{pPXF} \citep[for example][]{Cappellari2004,Cappellari2017}.

\begin{figure*}
	\includegraphics[width=2\columnwidth]{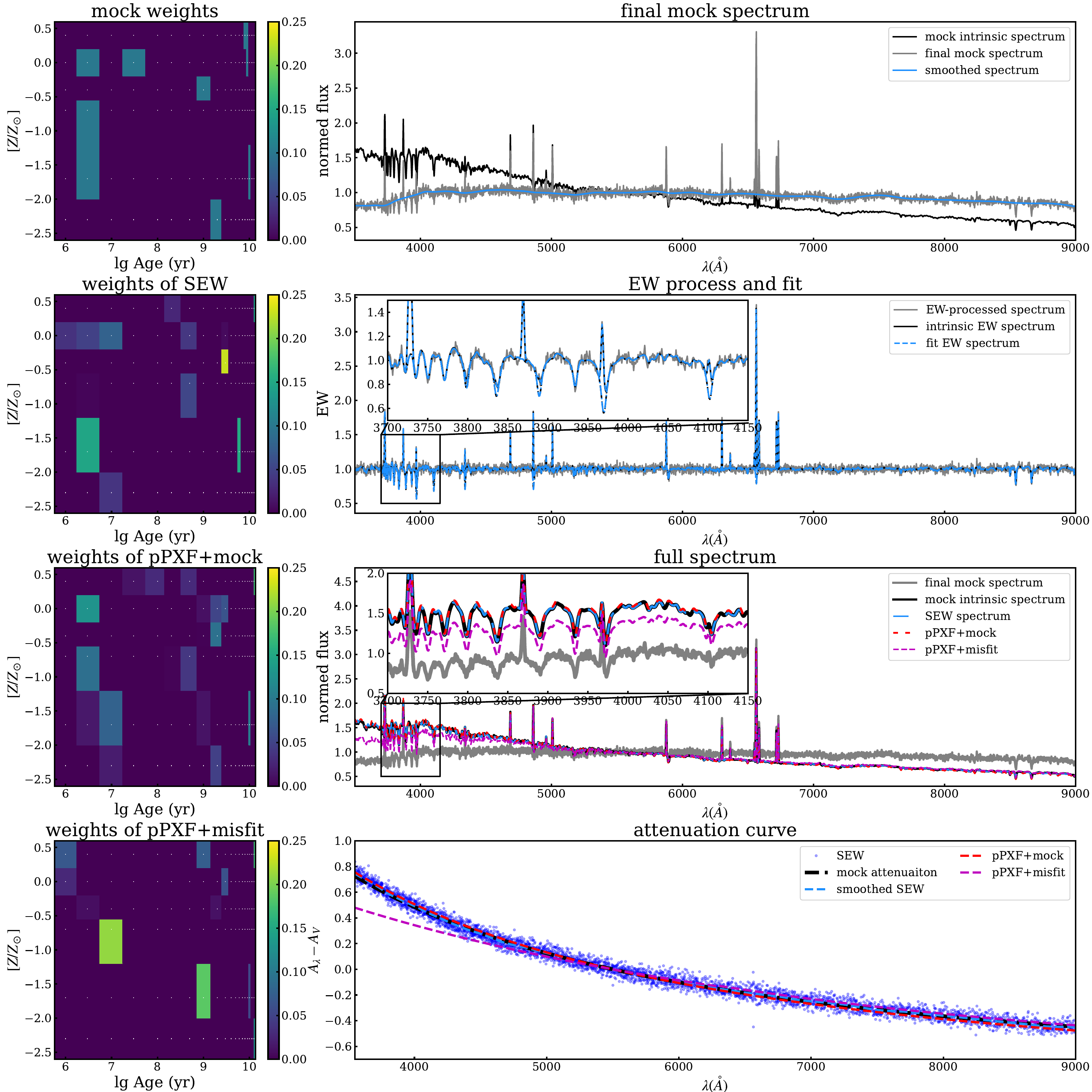}
    \caption{This figure illustrates the simple mock example spectrum used for our algorithm. Top-right: Spectral components showing the intrinsic spectrum (black) generated with the top-left SSPs distribution, attenuated mock spectrum with S/N=30 (grey, with the attenuation curve shown as black dashed lines in the bottom-right panel), and DPLS-smoothed spectrum (blue; Section \ref{sec:dpls}). Second-right: Equivalent width spectrum comparison between intrinsic (black), attenuated mock (grey), and SEW best-fit (blue dashed) with corresponding SSP weights in second-left panels. Third-right: Spectral reconstructions (combining Equation \ref{eq:csp} and the corresponding SSPs weights of each method) comparing intrinsic (black), attenuated mock (grey), SEW best-fit (blue dashed; second-left weights), \texttt{pPXF}+power-law (red dashed; third-left weights), and \texttt{pPXF}+misfit (magenta dashed; fourth-left weights). The inset highlights 3700-4150 \r{A} features. Bottom-right: Comparison of attenuation curves derived from different methods. The attenuation curve of SEW is derived by applying Equation \ref{eq:alam-av} to the best-fit spectrum of SEW and mock spectrum (blue and grey lines in third-right panel respectively), while the pPXF attenuation curve directly represents the parametric form derived from its fitting results. Line styles correspond to methodologies as detailed in the legend.}
    \label{fig:example} 
\end{figure*}

\section{Test with mock spectrum} \label{sec:mock}

\subsection{Generating mock spectra} \label{sec:mock_spec}

To evaluate the reliability of our method, we generate a series of mock spectra. The stellar population template is derived from BC03 \citep{Bruzual2003} and utilises the Chabrier initial mass function \citep{Chabrier2003}. We select 120 Simple Stellar Populations (SSPs) across 20 distinct stellar ages (ranging from 1 Myr to 13 Gyr) and 6 different metallicities (spanning from $0.005 Z_\odot$ to $2.5 Z_\odot$), covering wavelengths from approximately 3300 \r{A} to 9000 \r{A}, with a spectral sampling interval of $\Delta\lg \lambda = 10^{-4}$ (corresponding to the typical for SDSS processed data) and a full width at half maximum (FWHM) of 2.76 \r{A}. For comparative analysis, we randomly select 10 templates from these 120 SSPs, normalize them at the V band, and assign an equal weight of 0.1 to each stellar template. 

The intrinsic mock spectrum are attenuated using a power-law attenuation curve described by:
\begin{equation}
A(\lambda) = E(B-V) \frac{\lambda^{-\beta}}{4450 \, \text{\AA} ^{-\beta} - 5500 \, \text{\AA} ^{-\beta}},
\end{equation}
where $E(B-V)$ represents the reddening between the B band and V band, and $\beta$ is the power-law slope. 

When generating mock spectrum, we also incorporate emission lines and kinematic information as illustrated in Figure \ref{fig:example} to demonstrate the reliability of SEW method in spectral smoothing and equivalent width calculations. However, since emission line and kinematic measurements are not the focus of this study, detailed settings for these components will not be elaborated here; anyone interested can refer to the example file provided in our code repository for further information.

The top row of Figure \ref{fig:example} displays a spectrum generated through the process described above. In this panel, the left portion illustrates the composition of the stellar population, while the intrinsic spectrum is represented by the black line in the right portion. This mock spectrum is characterised by a luminosity-weighted characteristic logarithmic stellar age $\log \text{age}_L/yr$ of $8.5$ and a characteristic metallicity $[Z/Z\odot]$ of $-0.6$, as detailed in the 2$^{nd}$ row of Table \ref{tab:example}. To further refine the mock, we apply an attenuation curve with parameters $E(B-V) = 0.3$ and $\beta = 1.3$. The final mock spectrum, incorporating kinematics, attenuation, and noise with a signal-to-noise ratio (S/N) of $30$, are shown as grey lines.

This mock spectrum is analysed in Section \ref{sec:simple_mock} using our method to test its performance under controlled conditions. Additionally, to account for potential calibration biases and spectral stitching deviations, we generate two more mock spectra for further testing in Section \ref{sec:extra_mocks}. In this paper, we focus our comparative analysis on benchmarking \texttt{pPXF-SEW} against the conventional \texttt{pPXF} method to evaluate the robustness against extinction and calibration biases of SEW method.  Readers interested in broader comparisons with other stellar population synthesis algorithms may consult \citet{Woo2024} for comprehensive testing results. All mocks and corresponding results presented in this section can be reproduced using example files provided with \texttt{pPXF-SEW}.

\begin{table}
    \centering
    \caption{The bestfit $\log \text{age}_L/yr$ and $[Z/Z_\odot]$ of the simple example (Section \ref{sec:simple_mock}), calibration bias example and stitching artefacts example (Section \ref{sec:extra_mocks}). Here, the errors are estimated by non-negative least squares of the covariance matrix.}
    \label{tab:example}
    
    \begin{tabular}{@{}llcc@{}}
    \toprule
    Example & Method & $\log \text{age}_L/yr$ & $[Z/Z_\odot]$ \\
    \midrule
    \multicolumn{2}{c}{Intrintic}    & 8.5 & -0.6 \\
    \midrule
    \multirow{3}{*}{Simple Example} 
        & SEW                    & $8.5  \pm 0.18$ & $-0.65 \pm 0.23$ \\
        & \texttt{pPXF}+mock     & $8.53 \pm 0.18$ & $-0.54 \pm 0.57$ \\
        & \texttt{pPXF}+misfit   & $8.69 \pm 0.10$ & $-0.73 \pm 0.15$ \\
    \midrule
    \multirow{3}{*}{Calibration Bias} 
        & SEW                    & $8.51 \pm 0.17$ & $-0.66 \pm 0.18$ \\
        & \texttt{pPXF}+curve    & $7.98 \pm 0.04$ & $-0.11 \pm 0.05$ \\
        & \texttt{pPXF}+poly     & $8.11 \pm 0.02$ & $-0.13 \pm 0.03$ \\
    \midrule
    \multirow{3}{*}{stitching artefacts} 
        & SEW                    & $8.49 \pm 0.14$ & $-0.64 \pm 0.18$ \\
        & \texttt{pPXF}+curve    & $8.78 \pm 0.16$ & $-1.23 \pm 0.16$ \\
        & \texttt{pPXF}+poly     & $8.57 \pm 0.04$ & $-1.04 \pm 0.07$ \\
    \bottomrule
    \end{tabular}
\end{table}

\subsection{Bias-free attenuation with SEW} \label{sec:simple_mock}

In the top row of Figure \ref{fig:example}, we present the smoothed spectrum, represented by the blue solid line, obtained using the DPLS method described in Section \ref{sec:dpls}. The $\sigma$ is $10^{-8}$ (more detailed discuss in Section \ref{sec:sigma}) in equation \ref{eq:wf} and masking the emission lines by the weights $\bm{w}$. By combining this with the galaxy spectrum, we derive the EW spectrum, as indicated by the grey solid line in the right portion of the $2^{nd}$ row of Figure \ref{fig:example}. Using the SEW method outlined in Section \ref{sec:method}, we obtain the best-fit results for the equivalent width spectrum.

The weight fractions are shown in the left portion of the $2^{nd}$ row. The best-fit equivalent width spectrum corresponding to the weight is plotted in blue in the corresponding right panel. Here, the dashed line indicates the spectrum incorporating emission lines, while the dash-dotted line shows the spectrum without emission lines. A solid grey line marks the intrinsic equivalent width spectrum for comparison. To highlight detailed features, we zoom in on the region from 3700 to 4150 \r{A} (associated with D4000) in the inset box, where abundant emission and absorption lines are visible. The best-fit parameters, $\log \text{age}_L/yr = 8.51$ and $[Z/Z_\odot] = -0.66$, are tabulated in Row 3 of Table \ref{tab:example}. The linearisation is what allows us to obtain the corresponding weights for each SSP. Although small discrepancies between fitted and mock weight fractions arise from numerical effects and systematic uncertainties, their global distributions and spectral profiles show good agreement. Statistical characteristics (including age and metallicity estimates) of these equivalent width spectra match closely with those derived from their intrinsic counterparts. Observed minor deviations likely stem from well-documented age-metallicity degeneracies influencing such measurements.

In the right panel of the third row in Figure \ref{fig:example}, we show the full-spectrum reconstruction derived from fitting results using \texttt{pPXF-SEW} with Equation \ref{eq:csp}. The black line corresponds to the intrinsic mock spectrum, while the blue line shows the SEW method fitted spectrum. For comparison, we additionally performed stellar population synthesis on this mock spectrum using standard \texttt{pPXF}. During this analysis, we applied both our mock attenuation curve (power-law form) and a misfit curve, the Calzetti law curve \citep{Calzetti2000}. The resulting spectra from these two approaches appear as red and magenta dashed lines in the same panel. The derived parameters $\log \text{age}_L/yr = 8.53$ and $[Z/Z_\odot]=-0.54$ for the mock attenuation curve compare with $\log \text{age}_L/yr=8.69$ and $[Z/Z_\odot]=-0.73$ for the misfit curve (see Rows 4-5 in Table \ref{tab:example}). Corresponding weight fractions appear in left panels 3-4 of Figure \ref{fig:example}. An inset magnifies key features between 3700-4150 \r{A} for detailed inspection. The reconstruction of SEW method and \texttt{pPXF} results with mock attenuation curve closely match the intrinsic mock spectrum. Conversely, fits using a misfit curve display marked deviations from ground truth values across multiple spectral features.

\begin{figure*}
    \centering
    \includegraphics[width=2\columnwidth]{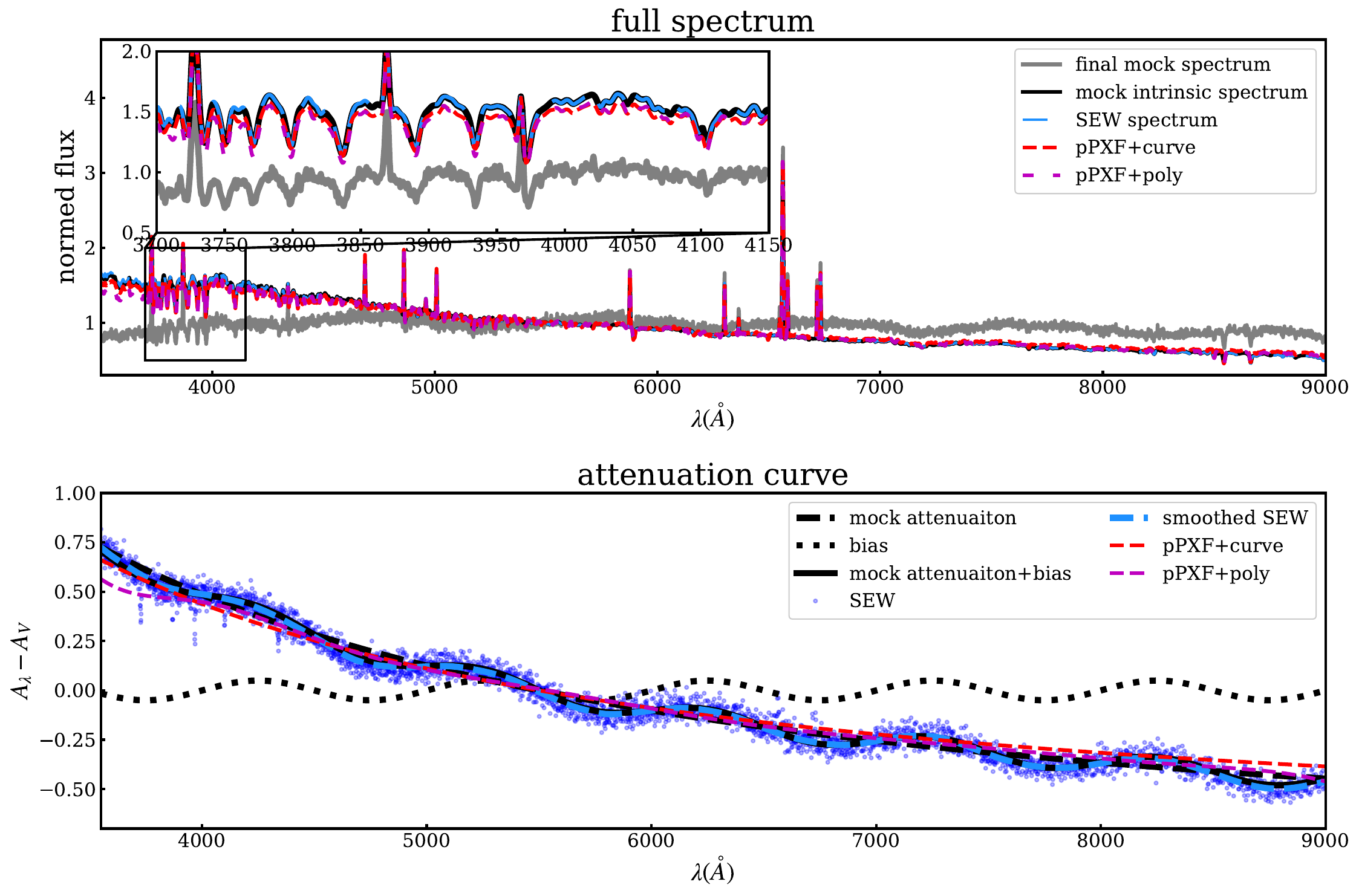}        
     \caption{Comparison of different spectral processing methods in the presence of flux calibration bias. The top panel shows the full spectrum, including the input spectrum (grey), the mock intrinsic spectrum (black), the EW fit spectrum (bule), and the results from \texttt{pPXF}+mock curve (red) and \texttt{pPXF}+misfit (magenta). The inset zooms in on a specific wavelength range (3700-4150 \r{A}) to highlight the differences. The bottom panel displays the attenuation curve, showing the mock attenuation (dashed line), the calibration bias(dotted line) , the mock attenuation with bias (solid line), the EW fit results (blue dots and line), and the results from \texttt{pPXF} + mock curve (red) and \texttt{pPXF}+misfit (magenta).}
    \label{fig:calbias}
\end{figure*}

\begin{figure*}
    \centering
    \includegraphics[width=2\columnwidth]{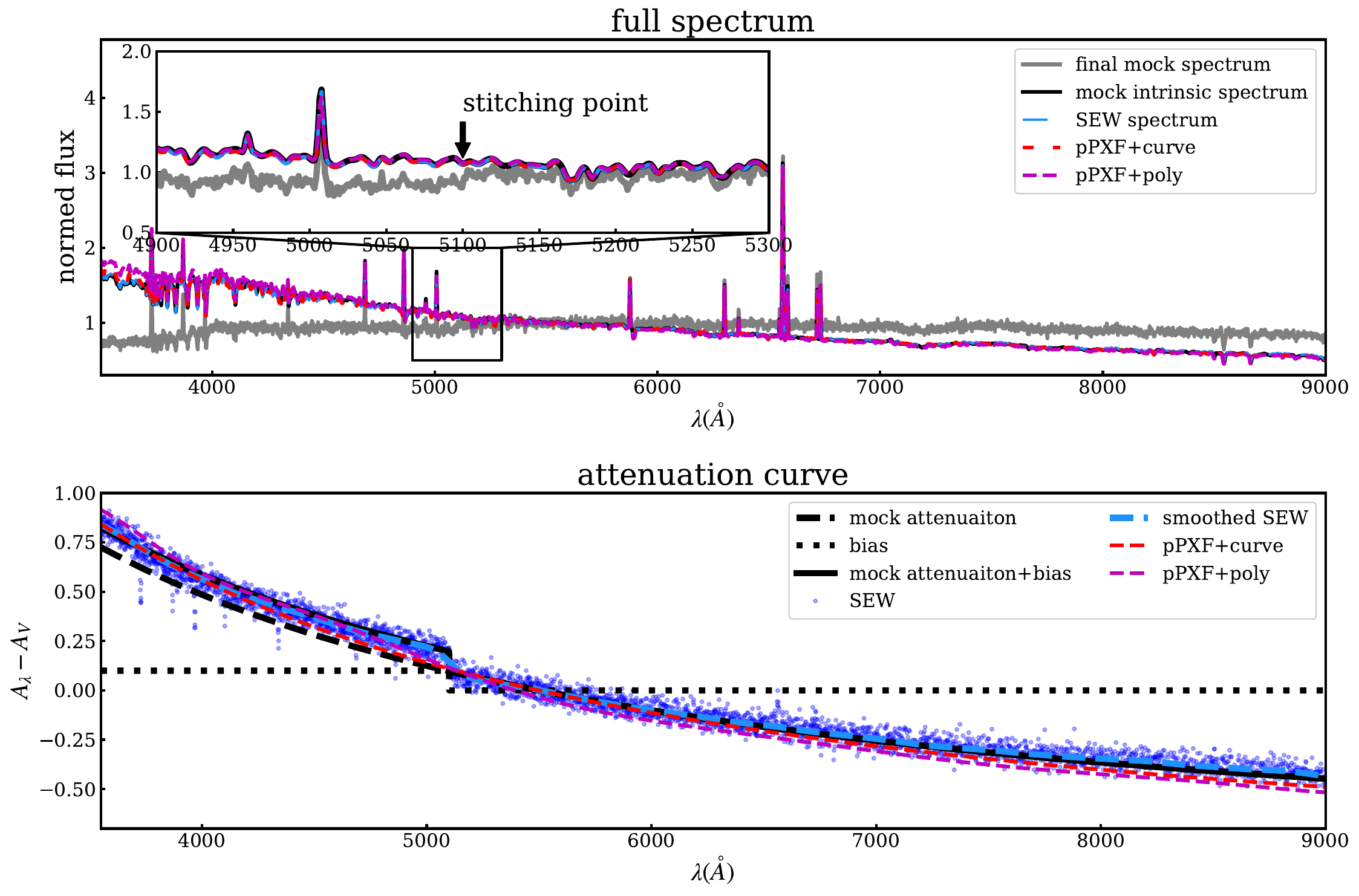}       
    \caption{Same with Fig. \ref{fig:calbias} but for spectral stitching artefacts.}
    \label{fig:stibias}
    \end{figure*}

The attenuation curve derived from our method using Equation \ref{eq:alam-av} is shown as blue dots in the lower right panel of Figure \ref{fig:example}, with the smoothed curve indicated by a blue dashed line. This panel compares the other two approaches: the red dashed line represents the power-law fit from \texttt{pPXF}, while the magenta dashed line corresponds to the misfit curve. The intrinsic attenuation curve of the mock spectra is shown in black for reference. Analysis of both spectral and attenuation results demonstrates close consistency between \texttt{pPXF} fits using "true" attenuation curve and those obtained through our method, as further corroborated by the statistical results in Appendix \ref{app:curve}. However, significant discrepancies emerge when applying the misfit curve, particularly evident in both spectral fitting and attenuation curve reconstruction compared to mock data. 

This underscores why classical stellar population synthesis requires precise dust law specification and how inappropriate dust law selection can introduce systematic biases in stellar population analysis - a critical concern given that the ``true" attenuation curves are generally not well constrained a priori, or whether empirical attenuation curves can encompass all the complex attenuation scenarios. The SEW method eliminates this requirement by self-consistently generating attenuation curves as part of its output. This intrinsic capability effectively avoids selection-induced biases, providing more robust stellar population characterisation compared to conventional approaches requiring predefined dust laws.

\subsection{Flux bias robustness of SEW} \label{sec:extra_mocks}

Observational spectral profiles exhibit systematic distortions arising not only from astrophysical attenuation but also from flux calibration uncertainties. These systematic effects typically manifest as either continuum scaling discrepancies or inter-segment discontinuities, primarily originating from two sources: imperfections in flux calibration procedures and artefacts introduced during spectral stitching processes. 

Flux calibration constitutes the critical conversion of raw detector counts into physically meaningful flux units, which depends fundamentally on reference standard stars with well-characterised spectral energy distributions. However, systematic errors (termed flux calibration bias) may emerge from three principal sources: inaccuracies in reference star fluxes, variable atmospheric transmission during observations, and unaccounted instrumental response characteristics.

Spectral stitching refers to the technical challenge of assembling discrete spectral segments-obtained through different instrumental configurations or wavelength regimes-into a continuous spectrum. This process risks introducing stitching artefacts due to inter-band calibration mismatches, variations in instrument response functions, and algorithmic limitations in data reduction pipelines.

This investigation employs two targeted mocks to evaluate the resilience of SEW methodology against distinct calibration bias, the flux calibration bias and the stitching artefacts. Both mock datasets maintain identical stellar population parameters, dust attenuation properties, and kinematic configurations to those defined in Section \ref{sec:simple_mock}. Our comparative framework implements: (1) \texttt{pPXF} analysis incorporating physically-motivated attenuation curves (\texttt{pPXF}+curve), simulating ideal attenuation treatment without accounting for calibration systematic bias; (2) \texttt{pPXF} mitigation using a tenth-order multiplicative polynomial (\texttt{pPXF}+poly), representing empirical calibration correction lacking physical attenuation modelling.

For flux calibration bias, we introduce periodic perturbations through sinusoidal modulation (1000 \r{A} wavelength periodicity, 0.05 mag amplitude; see black dashed curve in Figure \ref{fig:calbias}, lower panel). Spectral stitching artefacts are modelled via a localized 0.1 mag flux depression --applied exclusively to the blue segment ($\lambda < 5100$ \r{A}) at the merging boundary (black dashed line in Figure \ref{fig:stibias}, lower panel). The introduced bias settings are informed by potential artifacts encountered in real observational scenarios: Periodic gain variations caused by circuit design limitations may produce the flux calibration bias described in our setup, while calibration discrepancies between red and blue CCD detectors could generate spectral stitching artifacts. To accentuate the systematic effects of calibration biases, we deliberately configured these artificial perturbations with heightened amplitude and non-smooth variations.

Figures \ref{fig:calbias} and \ref{fig:stibias} present our recovery results using standardized panel layouts: upper panels compare reconstructed unattenuated spectra (colour-coded by solution type), while lower panels display corresponding recovered attenuation curves. Quantitative stellar population parameters ($\log \text{age}_L/yr$ and $[Z/Z_\odot]$) are systematically tabulated in Table \ref{tab:example}.

Figures \ref{fig:calbias} and \ref{fig:stibias} demonstrate the robustness of SEW method against both flux calibration and stitching artefacts through spectral reconstructions that closely match the intrinsic spectrum. This performance indicates dual capability of SEW to simultaneously mitigate large-scale systematic distortions (e.g., flux calibration bias) and localized discontinuities (e.g., stitching artefacts).

Conventional SPS methods lacking explicit bias correction mechanisms (\texttt{pPXF}+curve case) exhibit significant deviations from true mock values, as demonstrated by discrepancies in recovered $\log \text{age}_L/yr$ and $[Z/Z_\odot]$ parameters, along with mismatches in reconstructed intrinsic spectra and evolutionary curves (red lines in Figures \ref{fig:calbias} and \ref{fig:stibias}) when compared with SEW results and ground truth simulations.

In contrast, parametric bias calibration approaches (\texttt{pPXF}+poly case) fail to achieve satisfactory bias correction even when employing tenth-order polynomials ($\log \text{age}_L/yr$ and $[Z/Z_\odot]$ parameters of \texttt{pPXF}+poly in Table \ref{tab:example} and magenta lines in Figures \ref{fig:calbias} and \ref{fig:stibias}). Furthermore, this nonlinear optimization process substantially increases computational costs during SPS fitting - in our test case, the \texttt{pPXF}+poly implementation exceeded the runtime of the SEW method.

Notably, SEW-derived parameters show remarkable consistency with the true values of mock spectrum. The reconstructed spectral variations encode information about both dust attenuation properties and residual flux systematics. This stands in contrast to classical stellar population synthesis methods employing standard attenuation laws, which inherently fail to account for observational biases like unanticipated stitching artefacts. Such limitations ultimately compromise their ability to recover accurate galaxy properties, particularly star formation histories and chemical enrichment patterns, as quantified in Table \ref{tab:example}.

\begin{figure*}
    \centering
    \includegraphics[width=2\columnwidth]{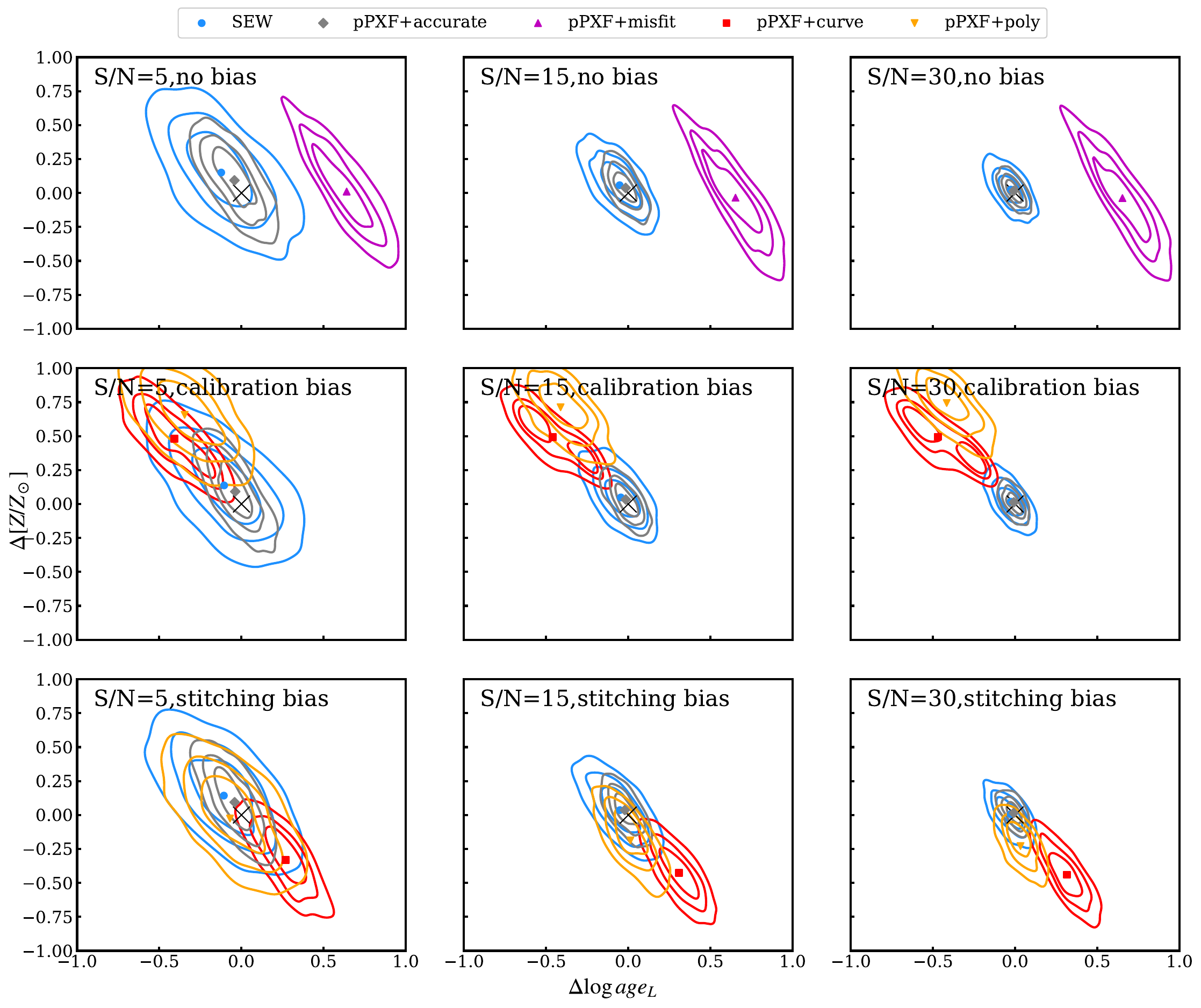}      
    \caption{The deviations of age-metallicity (kernel density estimate) KDE distribution for different methods and various S/N and flux bias settings. Each subplot displays a specific test condition, labeled in the upper left corner of the plot. Different methodologies are colour-coded in the panel legend, with reference zero-points marked by 'X' symbols.}
    \label{fig:age_metal} 
\end{figure*}

An important caveat emerges from our analysis: The global spectral variations recovered through SEW represent a superposition of dust attenuation signatures and residual flux systematics. Our current implementation does not permit direct decomposition of these intertwined components. Future studies could explore parametric modelling or non-parametric separation techniques to isolate pure attenuation curves, an extension beyond this work scope that nevertheless highlights promising research directions.

Despite this limitation, the SEW framework marks substantial progress in observational spectral analysis by maintaining parameter recovery accuracy even when confronting complex observational challenges. This advancement holds particular significance for next-generation spectroscopic surveys requiring robust analysis pipelines capable of handling heterogeneous datasets.

\begin{table*}
    \centering
    \begin{tabular}{cc|cccccc}
        \hline
        \multirow{2}{*}{Bias type} & 
        \multirow{2}{*}{Method} & 
        \multicolumn{2}{c}{S/N=5} & 
        \multicolumn{2}{c}{S/N=15} & 
        \multicolumn{2}{c}{S/N=30} \\
        \cline{3-8}
        & & $\Delta \log \text{age}_L$ & $\Delta[Z/Z_\odot]$ & $\Delta \log \text{age}_L$ & $\Delta[Z/Z_\odot]$ & $\Delta \log \text{age}_L$ & $\Delta[Z/Z_\odot]$ \\
        \hline
        \multirow{3}{*}{No bias} 
        & SEW & $-0.12 \pm 0.27$ & $0.14 \pm 0.36$ & $-0.05 \pm 0.15$ & $0.06 \pm 0.22$ & $-0.03 \pm 0.1$ & $0.03 \pm 0.14$ \\
        \cline{2-8}
        & pPXF+accurate & $-0.04 \pm 0.15$ & $0.08 \pm 0.27$ & $-0.02 \pm 0.09$ & $0.04 \pm 0.16$ & $-0.01 \pm 0.06$ & $0.01 \pm 0.1$ \\
        \cline{2-8}
        & pPXF+misfit & $0.64 \pm 0.2$ & $0.01 \pm 0.34$ & $0.65 \pm 0.18$ & $-0.03 \pm 0.34$ & $0.65 \pm 0.18$ & $-0.04 \pm 0.35$ \\
        \hline
        \multirow{4}{*}{Calibration bias} 
        & SEW & $-0.09 \pm 0.27$ & $0.13 \pm 0.36$ & $-0.05 \pm 0.14$ & $0.06 \pm 0.2$ & $-0.03 \pm 0.1$ & $0.04 \pm 0.15$ \\
        \cline{2-8}
        & pPXF+accurate & $-0.05 \pm 0.15$ & $0.11 \pm 0.27$ & $-0.02 \pm 0.09$ & $0.04 \pm 0.15$ & $-0.01 \pm 0.06$ & $0.02 \pm 0.11$ \\
        \cline{2-8}
        & pPXF+curve & $-0.4 \pm 0.19$ & $0.48 \pm 0.27$ & $-0.46 \pm 0.19$ & $0.5 \pm 0.22$ & $-0.47 \pm 0.19$ & $0.5 \pm 0.21$ \\
        \cline{2-8}
        & pPXF+poly & $-0.35 \pm 0.24$ & $0.66 \pm 0.28$ & $-0.41 \pm 0.18$ & $0.71 \pm 0.23$ & $-0.42 \pm 0.17$ & $0.74 \pm 0.22$ \\
        \hline
        \multirow{4}{*}{Stitching bias} 
        & SEW & $-0.11 \pm 0.27$ & $0.13 \pm 0.35$ & $-0.05 \pm 0.14$ & $0.05 \pm 0.21$ & $-0.03 \pm 0.1$ & $0.02 \pm 0.14$ \\
        \cline{2-8}
        & pPXF+accurate & $-0.05 \pm 0.15$ & $0.1 \pm 0.27$ & $-0.02 \pm 0.09$ & $0.04 \pm 0.16$ & $-0.01 \pm 0.06$ & $0.02 \pm 0.1$ \\
        \cline{2-8}
        & pPXF+curve & $0.27 \pm 0.16$ & $-0.33 \pm 0.27$ & $0.31 \pm 0.13$ & $-0.43 \pm 0.22$ & $0.31 \pm 0.12$ & $-0.44 \pm 0.21$ \\
        \cline{2-8}
        & pPXF+poly & $-0.07 \pm 0.25$ & $-0.03 \pm 0.33$ & $0.01 \pm 0.13$ & $-0.19 \pm 0.22$ & $0.03 \pm 0.10$ & $-0.23 \pm 0.17$ \\
        \hline
    \end{tabular}
    \caption{The statistical results for different bias types and fitting methods at various S/N levels (corresponding to Figure \ref{fig:age_metal}). The results are formatted as $\overline{\Delta \log \text{age}_L} \pm \sigma_{\Delta \log \text{age}_L}$ and $\overline{\Delta[Z/Z_\odot]} \pm \sigma_{\Delta[Z/Z_\odot]}$.}
    \label{tab:sta_age_Z}
\end{table*}

\subsection{Statistical Results}  \label{sec:stat}

To quantitatively evaluate the SEW method, we synthesize 1,000 mock spectra following the procedures detailed in Section \ref{sec:mock_spec}. We conduct systematic evaluations of the method reliability across varying signal-to-noise ratios (S/N = 5, 15, 30) under three distinct flux calibration scenarios: unbiased flux calibration, calibration-induced bias, and spectral stitching artefacts (as defined in Sections \ref{sec:simple_mock} and \ref{sec:extra_mocks}). The attenuation model employs a power law parameterized with$E(B-V)$= 0.3 and $\beta = 1.3$. For each combination of flux calibration condition and S/N level, we apply identical fitting methodologies to those described in the aforementioned sections. In biased conditions (calibration bias and stitching artefacts), we additionally implement a \texttt{pPXF}-based approach incorporating ``accurate curve with systematic error correction'' to assess potential improvement.

Figure \ref{fig:age_metal} displays the distributions of fitting deviations from true values $(\Delta \log \text{age}_L, \Delta [Z/Z_\odot])$, with each subplot corresponding to specific test conditions annotated in the top-left corner (for direct comparisons between the true values and fitted values, see Appendix \ref{app:age_metal_dist}). Different methodologies are colour-coded in the panel legend, with reference zero-points marked by 'X' symbols. Table \ref{tab:sta_age_Z} quantifies these systematic offsets and their associated dispersions $(\overline{\Delta \log \text{age}_L}, \sigma_{\Delta \log \text{age}_L}),(\overline{\Delta[Z/Z_\odot]},\sigma_{\Delta[Z/Z_\odot]})$ across various bias types and S/N levels.

Our analysis reveals comparable performance between the SEW method and \texttt{pPXF} with properly calibrated attenuation curves, both showing minor deviations within statistical uncertainty ranges (Table \ref{tab:sta_age_Z}) in both unbiased calibration and biased calibration scenarios. The observed discrepancies likely originate from noise-induced spectral misidentification. In particular, misidentification of error signals as metal absorption lines leads to overestimation of metallicity at low metallicity with low S/N (bottom panel of Figure \ref{fig:mock_age_metal})

The results of SEW method, however, show greater dispersion compared to those from \texttt{pPXF} using the accurate calibrated curve, showing an approximately 1.5-fold increase in dispersion\footnote{This dispersion enhancement shows an inverse correlation with the penalty scale factor $\sigma$, as detailed in Section \ref{sec:sigma}.}. This increased dispersion originates from our strategic removal of continuum spectral components during measurement, which inevitably sacrifices some photometric information. We argue this intentional information loss constitutes a justified compromise: By eliminating continuum contributions, we effectively mitigate systematic uncertainties inherent to attenuation curve selection and calibration biases that typically plague traditional methods.

Comparative analyses using \texttt{pPXF} with mismatched attenuation laws (magenta contours in the top panels of Figures \ref{fig:age_metal} and \ref{fig:mock_age_metal}) reveal significantly larger systematic deviations. These findings are consistent with the demonstration in Section \ref{sec:simple_mock}, reinforcing the critical dependence on attenuation curve selection inherent to conventional stellar population synthesis methods. Our methodology inherently eliminates this selection bias.

Conversely, spectra exhibiting calibration systematics demonstrate substantial fitting deviations even when employing appropriate attenuation curves, provided that calibration offsets remain unaccounted for (red contours in lower panels of Figures \ref{fig:age_metal} and \ref{fig:mock_age_metal}). Of course, higher calibration precision naturally yields smaller residual biases, eventually degrading to a bias-free case.  While standard practice addresses such systematics through high-order polynomial corrections, our testing demonstrates that even a 10th-degree polynomial fails to fully encapsulate calibration biases (orange contours in corresponding panels) in our cases, resulting in residual deviations exceeding acceptable thresholds. This is because polynomial corrections need to take into account both the calibration bias and the existence of extinction. This limitation stems from intrinsic challenges in multiplicative polynomial calibration: the arbitrary selection of polynomial order introduces artificial effects, where insufficient polynomial degrees fail to resolve intermediate-scale spectral deviations, while excessive degrees induce non-linear parameter degeneracies (obscuring physical parameters amidst numerous polynomial coefficients) and escalate computational demands through prolonged optimisation processes.

The fitting results without any prior calibration bias correction or appropriate attenuation curve selection exhibit significant deviations from the intrinsic spectrum. We emphasize that these deviations are specific to the predefined mock scenarios in this paper. While these deviations are specific to our predefined mock scenarios, we emphasize that the adopted about $5\%$ flux bias serves as a deliberately strong assumption to amplify the potential impact of systematic errors for illustrative purposes. In real observations, it is difficult to precisely assess various sources of bias. The systematic biases in stellar population synthesis results caused by such flux biases and attenuation curve choice bias are also challenging to estimate. Although our simplified bias model highlights first-order effects, we caution that real systematic uncertainties likely arise from compounded effects, including degeneracies between stellar population parameters, attenuation curve flexibility, and wavelength-dependent flux calibration errors, whose collective behaviour requires more sophisticated error propagation frameworks. For instance, recently detected intermediate-scale structures in extinction curves (FWHM $\sim$ 100 \r{A}; \citealt{Massa2020,Zhang2024,Green2024}) highlights how insufficient polynomial order compromises measurement accuracy of spectral features.

In contrast,the results indicate that the SEW method is robust against complex attenuation and calibration biases, as the fitting results only shows minor deviations within the statistical error in age and metallicity. The enhanced robustness of our SEW approach enables non-parametric derivation of galaxy attenuation curves during stellar population synthesis (demonstrated in Appendix \ref{app:curve}). This capability facilitates detailed investigations into attenuation curve morphology, potentially revealing fine-scale structures that conventional methods might obscure.

\section{Discussion} \label{sec:discussion}

\subsection{the choisen of penalty scale factor \texorpdfstring{$\sigma$}{sigma}} \label{sec:sigma}

\begin{figure*}
    \centering
    \includegraphics[width=2\columnwidth]{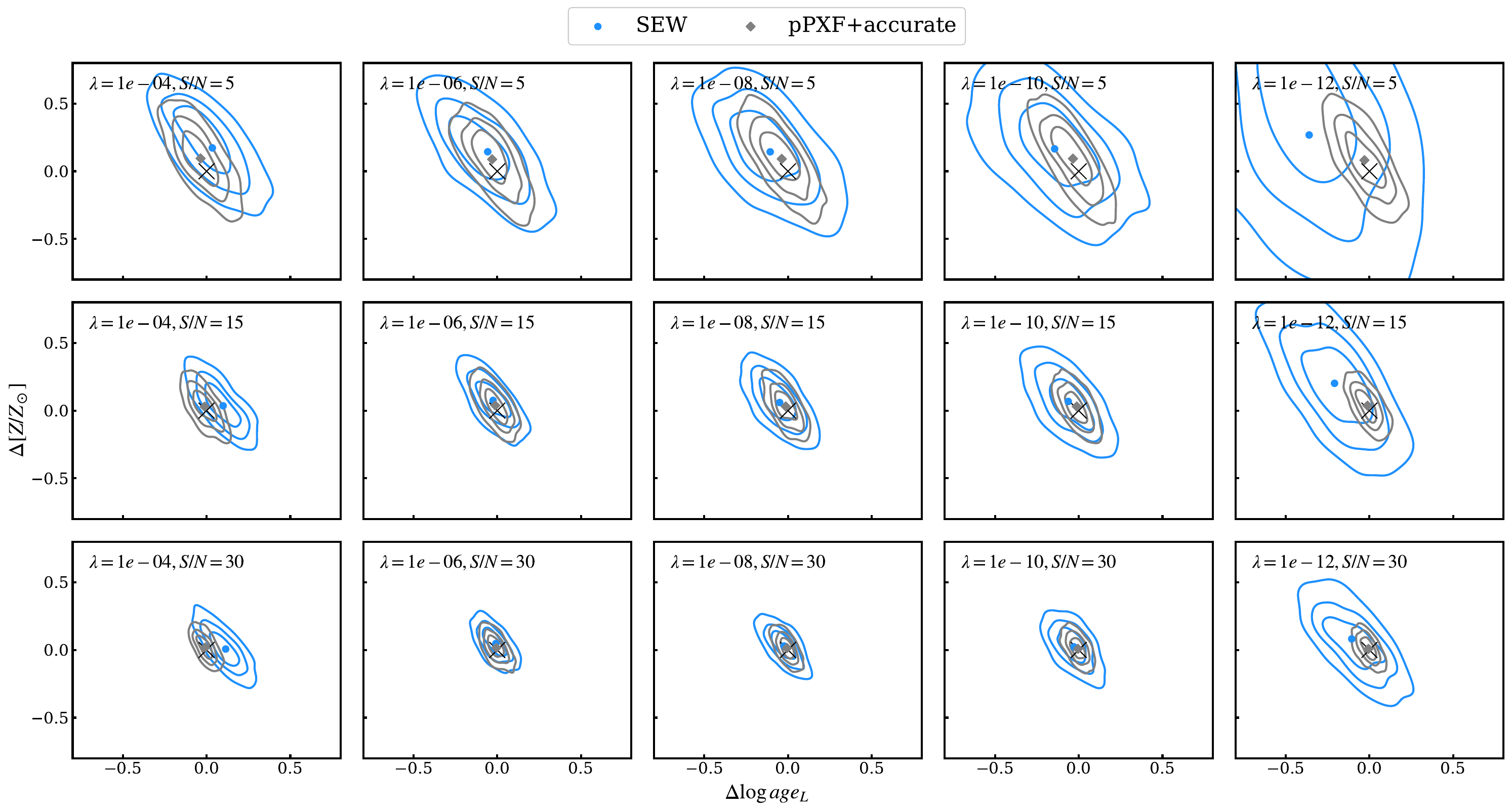}        
    \caption{The age-metallicity KDE distribution for different S/N and various $\lambda$ settings. Each subplot displays a specific test condition, labeled in the upper left corner of the plot. Each subplot involves two different methods: our method and \texttt{pPXF} with the ``true attenuation curve", each represented by a distinct colour as indicated in the legend at the top of the figure. The contour lines, from inner to outer, represent the first quartile, second quartile (median), and third quartile, respectively. The zero-point is marked with an 'X' in each subplot.}
    \label{fig:factor}
\end{figure*}

The penalty scale factor $\sigma$ in the DPLS method is a critical parameter that controls the smoothness of the continuum. A small $\sigma$ can lead to an absorption line signal deduction within the smoothed spectrum that is not clean, potentially retaining noise or residual features. Conversely, a large $\sigma$ can result in the loss of continuous spectrum information, causing over-smoothing and obscuring important details. This trade-off is fundamentally governed by the relationship between $\sigma$ and the characteristic scale of suppressed signal fluctuations, as derived from the scaling laws of the second-order difference penalty.

For discrete data sampled at intervals $\Delta x$, the second-order difference operator approximates the second derivative:
\begin{equation}
    \Delta^2 f_i  \sim \frac{f_{i+1} - 2f_i + f_{i-1}}{(\Delta x)^2}.
\end{equation}
We note that the scaling form of the $\Delta^2$ operator is $\frac{1}{(\Delta x)^2}$. In the penalised least squares framework ( Equation \ref{eq:wf}), the balance between the fitting term $\bm{WF'}$ and the penalty term $\sigma\bm{D_n}^{\top} \bm{D_n} \bm{F'}$ determines the effective smoothing scale. When these terms are balanced (i.e., $\sigma(\Delta^2)^2\sim 1$), the characteristic length scale $\Delta x$ in $\ln\lambda$-space satisfies $\Delta x \sim \sigma^{1/4}$, where $\sigma$ is the penalty scale factor. Translating this to linear wavelength space, the suppressed wavelength range $\Delta \lambda$ at a reference wavelength $\lambda_0$ is given by:
\begin{equation}
    \Delta \lambda \sim \lambda_0 \cdot \left(e^{\sigma^{1/4}} - 1\right) \sim \lambda_0 \cdot \sigma^{1/4} \quad (\text{for } \sigma^{1/4} \ll 1).
\end{equation}
For example, to suppress fluctuations of $\Delta \lambda = 20$\r{A} at $\lambda_0 = 5000$\r{A}, the required penalty scale factor is about 2.5 $\times 10^{-10}$. This relationship highlights how the penalty scale factor $\sigma$ controls the trade-off between resolving fine spectral features (small $\sigma$) and smoothing out noise or rapid variations (large $\sigma$).

In this section, we will systematically validate the choice of $\sigma$, aiming to find an optimal balance that preserves the integrity of the continuous spectrum while effectively removing noise and unwanted features. This will ensure that the absorption line characteristics are accurately represented in the processed data.  To explore this, we test how varying $\sigma$ affects the distribution of deviations between the fitted and true values of age and metallicity for the 1000 mock spectra. By analyzing these distributions, we can identify optimal $\sigma$ values that minimise the discrepancies between the fitted parameters and their known true values, thereby enhancing the accuracy and reliability of our method under different conditions. In this test, the $\sigma$ takes the values of $10^{-4}$, $10^{-6}$, $10^{-8}$, $10^{-10}$, $10^{-12}$ (corresponding to the characteristic scale $\Delta \lambda$ of about 500\r{A}, 160\r{A}, 50\r{A}, 16\r{A}, 5\r{A} at $\lambda_0 = 5000$\r{A}, respectively). Additionally, we apply an power-law attenuation curve with  $E(B-V)$  = 0.3 and $\beta$ = 1.3. 

The test results is shown in Figure \ref{fig:factor}. In this figure, we find that the performance of $\sigma$ in the range of $10^{-6}$ to $10^{-10}$ (corresponding to the characteristic scale $\Delta \lambda$ from about 160\r{A}, to 16\r{A} at $\lambda_0 = 5000$\r{A}) is not significantly different. As shown from the second to the fourth column in Figure \ref{fig:factor}, the mean of $(\Delta \log \text{age}_L, \Delta[Z/Z_\odot])$ does not vary substantially within this range. Slight deviations may be an amplification of intrinsic deviations in pPXF.  As shown in the figure, the results of our method have the same trend but slightly larger deviations compared to pPXF. However, when $\sigma$ is $10^{-4}$ ($\Delta \lambda \sim 500 $\r{A} at $\lambda_0 = 5000$\r{A}) and $10^{-12}$ ($\Delta \lambda \sim 5 $\r{A} at $\lambda_0 = 5000$\r{A}), the results of our method exhibit a much larger and off-trend deviation. 

The observed deviations at extreme penalty scale factor can be attributed to two physical limitations of the smoothing mechanism. For $\Delta \lambda \sim 5 $\r{A}, the smoothing scale becomes smaller than the typical FWHM of galactic emission/absorption lines, rendering it insufficient to suppress unresolved narrow features or noise. Conversely, for $\Delta \lambda \sim 500 $\r{A}, the excessive smoothing erases intermediate-scale spectral features that noise the equvilent width spectrum. The stable performance within $\sigma = 10^{-6}-10^{-10}$ ($\Delta \lambda ~ 160-16$\r{A}) aligns with the characteristic scales of galactic spectral components, balancing noise and line removal and feature preservation.

Within the range of $10^{-6}$ to $10^{-10}$, as $\sigma$ increases, the dispersion decreases. This is because a larger $\sigma$ tends to include less absorption line information, which can lead to a smoother continuum. However, a very large $\sigma$ might also result in the inclusion of residual continuous spectrum and calibration bias information in the equivalent width spectra. To balance these two effects, we recommend setting $\sigma$ to $10^{-8}
$, as it is uesd in Section \ref{sec:mock}. This choice strikes a balance between preserving the integrity of the continuous spectrum and effectively removing noise and unwanted features, ensuring that the absorption line characteristics are accurately represented in the processed data.

\subsection{Inconsistency of attenuation} \label{sec:inconsistency}

\begin{figure*}
    \centering
    \includegraphics[width=2\columnwidth]{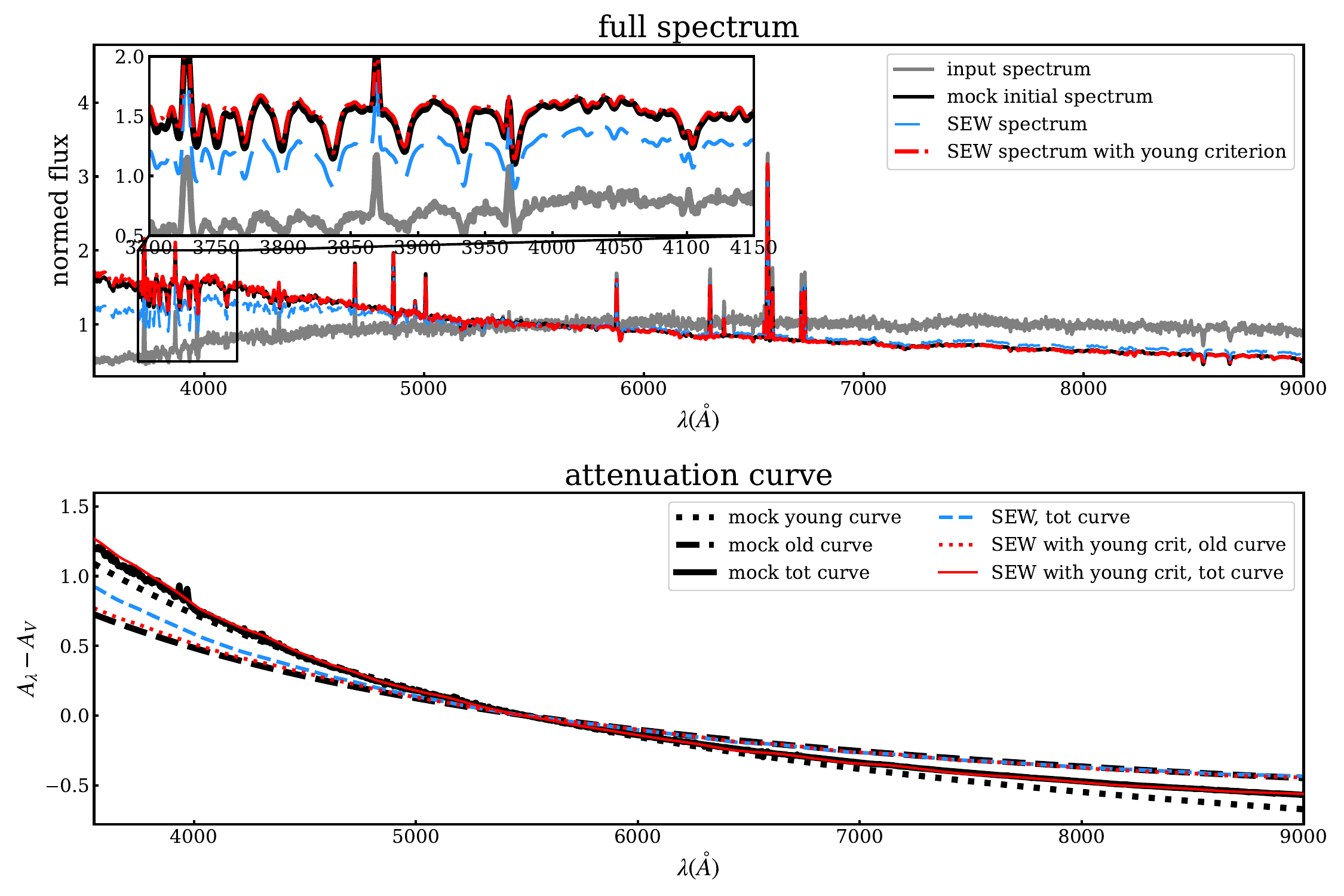}
    \caption{An example for multi-attenuation components. In this example, we performed a direct fit using our method (shown with blue lines), and incorporating an additional differential attenuation curve for the young stellar population (shown with red lines). The implication of the different lines is shown in the legend}
    \label{fig:multiatt}
\end{figure*}

Dust attenuation may exhibit more complex patterns that may vary across different stellar populations. Many studies reveal the attenuation difference between young and old stellar populations\citep{Calzetti1994,Mayya1996,CF00}. This difference is often attributed to a two-component dust model \citep[e.g.][]{Calzetti2000, CF00, Lu2022, Lu2023, Qin2024, Zhang2023}, wherein galaxies possess both a diffuse interstellar medium (ISM) and a clumpy birth-cloud component. Thus, the young stellar populations and nebular emission are attenuated by both the envelope of birth clouds and the diffuse ISM. While the stellar emission, especially that of old stellar populations, most of which is radiated outside of the birth clouds, is extincted only by the diffuse ISM component. This attenuation difference has been widely adopted in many studies \citep{Madau2014, Daddi2007, Peng2010}. 

In simplify, suppose that each stellar populations attenuation curve have a $\Delta \tau_j(\lambda)$ deviation from an average attenuation curves $\tau(\lambda)$:
\begin{equation}
\begin{aligned}
    EW&=\frac{\sum_j x_j f_{j}(\lambda)e^{-\tau(\lambda)-\Delta \tau_j(\lambda)}}{\sum_j x_j f'_{j}(\lambda)e^{-\tau(\lambda)-\Delta \tau_j(\lambda)}} \\
    &=\frac{\sum_j x_j f_{j}(\lambda)e^{-\Delta \tau_j(\lambda)}}{\sum_j x_j f'_{j}(\lambda)e^{-\Delta \tau_j(\lambda)}}.
\end{aligned}
\end{equation}
This average attenuation curves ensure most $\Delta \tau_j(\lambda) \sim 0$.

If all $\Delta \tau_j(\lambda) \sim 0$, this equation reduces to equation \ref{eq:ew}. This case (CASE I) is common in quenched galaxies. Quenched galaxies exhibit no star formation activity, with most of their stars being old. These stars are attenuated by the thin dust in the ISM that the attenuation consistent across different stellar populations.

Consider some stellar populations (donate as $j_1$) have $\Delta \tau_j(\lambda)$ that great larger than 1 or contribute very few flux ($x_{j_1} \sim 0$). In this case (CASE II), one can see that:
\begin{equation}
\begin{aligned}
    EW&=\frac{\sum_j x_j f_{j}(\lambda)+\sum_{j_1} x_{j_1} f_{j_1}(\lambda)e^{-\Delta \tau_{j_1}(\lambda)}}{\sum_j x_j f'_{j}(\lambda)+\sum_{j_1} x_{j_1} f'_{j_1}(\lambda)e^{-\Delta \tau_{j_1}(\lambda)}}\\
    &\approx \frac{\sum_j x_j f_{j}(\lambda)+0}{\sum_j x_j f'_{j}(\lambda)+0}
    \label{equ:case2}
\end{aligned}
\end{equation}
This equation also reduces to equation \ref{eq:ew}. In this case, these $j_1$ stellar populations barely contribute flux in optical wavelength. Therefore, these high attenuation stellar populations will be categorized as "absent" and assigned a weight of zero. As a result, the analysis of stellar populations will focus exclusively on those unaffected by significant attenuation. This case is often seen in metal-rich galaxies hosting giant molecular clouds and dust-rich HII regions. New-born stars in these regions are completely obscured by dust that almost all the optical flux is transferred to the infrared. Accurate assessment of these obscured populations necessitates multi-wavelength observations, particularly the infrared band.

It becomes complicated if the $\Delta \tau_j(\lambda)$ of some non-negligible populations (donate as $j_2$) fall between above two cases, i.e., $\Delta \tau_{j_2}(\lambda) \lesssim 1$. In this case, the equivalent width is written as:
\begin{equation}
    EW=\frac{\sum_j x_j f_{j}(\lambda)+\sum_{j_2} x_{j_2} f_{j_2}(\lambda)e^{-\Delta \tau_{j_2}(\lambda)}}{\sum_j x_j f'_{j}(\lambda)+\sum_{j_2} x_{j_2} f'_{j_2}(\lambda)e^{-\Delta \tau_{j_2}(\lambda)}}
\end{equation}
In this case (CASE III), it can be considered that the weight of $j_2$ population is modified by $\Delta \tau_{j_2}(\lambda)$. 

To mock this case, we implemented the stellar population distribution from Section \ref{sec:simple_mock}, separating populations at 0.1 Gyr with young stars contributing 40$\%$ V-band light and elevated reddening ($E(B-V)_{\text{young}}=0.45$ vs $E(B-V)_{\text{old}}=0.3$), maintaining identical $\beta=1.3$ powerlaw slopes for both components to produce $\sim$0.5 mag differential V-band attenuation.

We compared direct SEW fitting against a differential attenuation curve modelling (built-in parameters of pPXF, and we assume an additional power law differential attenuation curve) incorporating differential young population attenuation (Figure \ref{fig:multiatt}). Direct fitting (blue) shows spectral deviations below 4500\r{A} but matches old population attenuation above this threshold, while the complex model (red) better reproduces mock data through additional young star attenuation.

However, this does not necessarily mean that introducing the attenuation difference is a better fitting strategy. Fundamental uncertainties persist regarding the specific form of young population attenuation curves and which subpopulations require differential treatment. Additional parameters risk model degeneracy despite potential spectral improvements below 4500\r{A}. Crucially, direct fitting remains consistent with old population attenuation above 4500\r{A} (aligned with equation \ref{equ:case2}), showing minimal deviations without extra assumptions. We therefore recommend reserving complex models for scenarios with strong prior constraints on dust mechanisms, favouring direct fitting when physical parameters remain ambiguous.

\section{summary} \label{sec:summary}

Accurate determination of galactic physical properties from observed spectra remains a critical challenge in extragalactic astrophysics, particularly due to systematic uncertainties in dust attenuation modelling and flux calibration biases. In this work, we present SEW, a full-spectrum linear fitting method based on the "Equivalent Widths spectrum" (EWs), which eliminates the need for prior assumptions about dust attenuation curves while robustly addressing nonlinear degeneracies in traditional stellar population synthesis (SPS). By leveraging the intrinsic invariance of EWs to global spectral variations, our method employs the Discrete Penalised Least Squares (DPLS) technique to extract smoothed continua and derive attenuation curves as outputs rather than inputs. This approach linearises the fitting equations through matrix transformations, enabling efficient recovery of key parameters, including stellar age, metallicity, and dust attenuation, even under significant flux calibration errors or spectral stitching artefacts.

Rigorous validation using mock spectra across a wide signal-to-noise range (S/N = 5-30) demonstrates the method resilience. Systematic tests reveal minimal biases in recovered stellar population parameters and attenuation curves, even in scenarios with periodic flux miscalibrations or localized stitching discontinuities. The method's computational efficiency, requiring only 2-3 times the runtime of standard \texttt{pPXF} implementations, ensures practical applicability to large datasets. We further identify an optimal penalty scale factor range $\sigma \in [10^{-10},10^{-6}]$, corresponding to characteristic wavelength smoothing scales of $\Delta\lambda \approx$16--160\r{A} at a reference wavelength of 5000\r{A}, and recommend $\sigma=10^{-8}$ as a default for balancing feature preservation and noise suppression. 

While the method assumes uniform attenuation across stellar populations, we discuss limitations in cases of spatially varying dust distributions, emphasizing the need for complementary multi-wavelength data in such regimes. 

To facilitate adoption, we implement SEW as an open-source Python extension package (\texttt{pPXF-SEW}), which is available in \url{https://pypi.org/project/ppxf-sew/}, seamlessly integrated with the widely used pPXF framework. This work advances spectral fitting methodologies by decoupling attenuation modelling from stellar population analysis, offering a robust tool for next-generation galaxy surveys and high-precision extragalactic studies.

\section*{acknowledgments}

This work is partly supported by the National Key Research and Development Program (No. 2022YFA1602903, No. 2020SKA0110100, No. 2023YFB3002502), the National Natural Science Foundation of China (No. 12403016), the Postdoctoral Fellowship Program of CPSF (No. GZC20241514), and the science research grants from the China Manned Space project with NO.CMS-CSST-2021-A03, CMS-CSST-2021-B01. S.S. thanks research grants from the National Key Research and Development Program of China (No. 2022YFF0503402), Shanghai Academic/Technology Research Leader (22XD1404200) and the National Natural Science Foundation of China (No. 12141302). 

We appreciate the availability of \texttt{pPXF} and the detailed explanation of the licence by its authors, Michele Cappellari\footnote{\url{https://www-astro.physics.ox.ac.uk/~cappellari/}}. We promise that the \texttt{pPXF-SEW} package complies with licence of \texttt{pPXF}, and we suggest that any user of the \texttt{pPXF-SEW} package also acknowledge \texttt{pPXF}.

\section*{Data Availability}

All mocks and corresponding results presented in this paper can be reproduced using example files provided with \hyperlink{https://pypi.org/project/ppxf-sew/}{\texttt{pPXF-SEW}} in PYPI.



\appendix
\renewcommand\thefigure{\Alph{section}\arabic{figure}}   
\section{Attenuation Curve}\label{app:curve}
\setcounter{figure}{0} 

\begin{figure*}
    \centering
    \includegraphics[width=2\columnwidth]{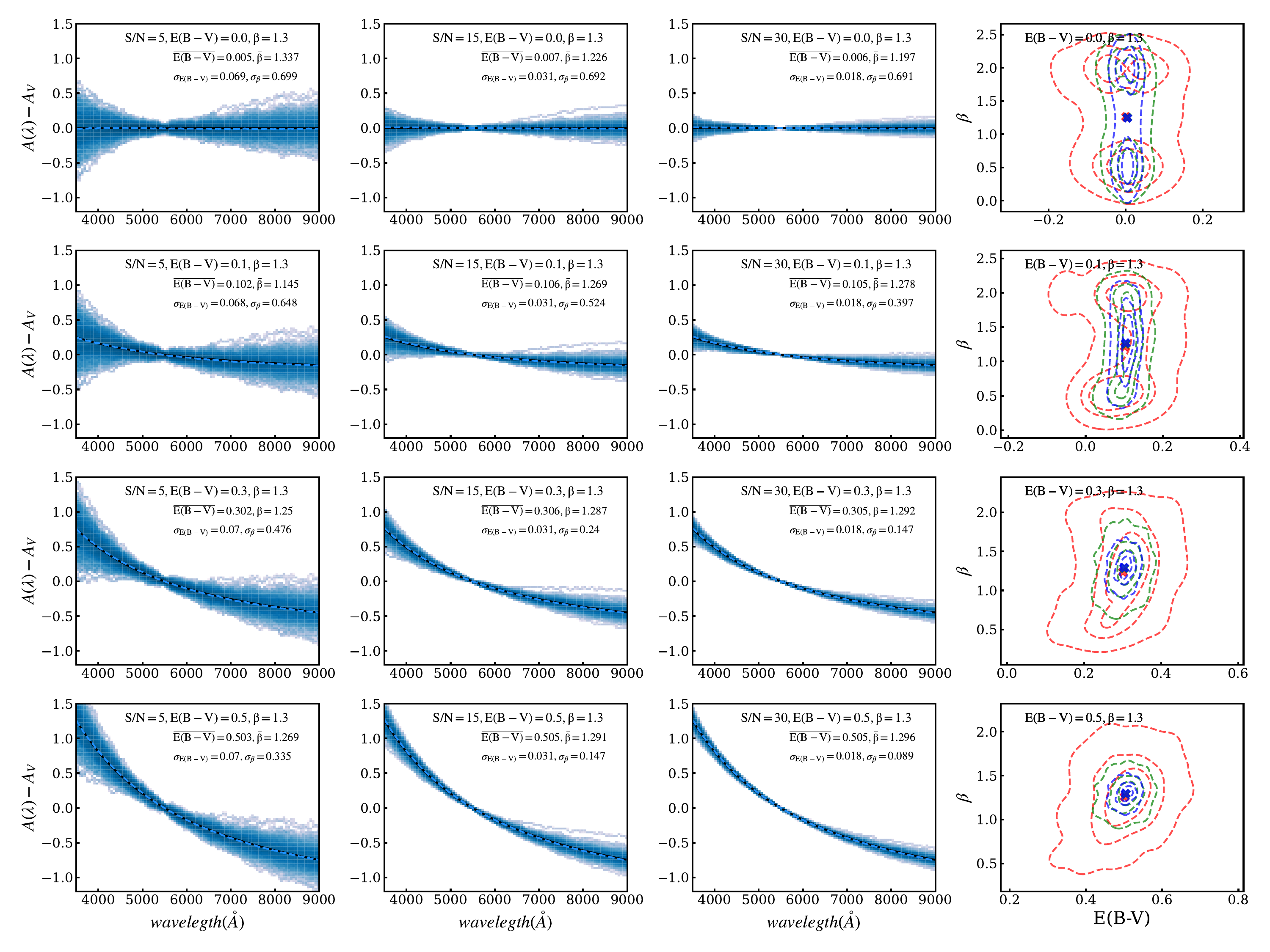}      
    \caption{This figure illustrates the attenuation curves obtained using the SEW method under different reddening ( $E(B-V)$ ) and signal-to-noise ratio (S/N) conditions. Each row corresponds to a specific  $E(B-V)$  value (0.0, 0.1, 0.3, and 0.5), and the left three columns show the attenuation curves for different S/N ratios (5, 15, and 30). In each panel of the left three columns, the density plot represents the distribution of 1000 mock fits, with the blue dashed line indicating the mean value and the black solid line representing the true value. The rightmost column displays the parameter distributions obtained by fitting the attenuation curves in the corresponding rows using a power law. Different colours in the rightmost column represent different S/N ratios: red for S/N = 5, magenta for S/N = 15, and blue for S/N = 30. The circle dot represents the intrinsic  $E(B-V)$  and $\beta$ and different colour ``X" mark represent the mean  $E(B-V)$  and $\beta$ at the corresponding S/N.}
    \label{fig:mock_curve_sn} 
\end{figure*}

This appendix validates attenuation curve recovery of the SEW method through controlled simulations using synthetic galaxy spectra. We systematically vary dust reddening $E(B-V) \in [0,0.1,0.3,0.5]$ (fixed $\beta=1.3$) and observational noise levels ($\mathrm{S/N}=5,15,30$), intentionally excluding calibration bias for direct benchmarking. Figure \ref{fig:mock_curve_sn} demonstrates consistent recovery of attenuation curves across all tested $E(B-V)$ values, with mean reconstructed curves (blue dashed lines) closely matching true values (black lines) despite increasing dispersion at lower S/N. The rightmost KDE plots reveal tight clustering around true parameters at S/N $\geq$15, while S/N=5 shows broader distributions. Notably, $E(B-V)$ dispersion remains constant across S/N levels due to combined uncertainties from stellar population fitting and random noise, whereas $\beta$ dispersion depends on $E(B-V)$ magnitude - weaker constraints emerge for lower $E(B-V)$ where attenuation curves flatten. These results confirm the robustness across diverse observational conditions, though precision decreases significantly at S/N=5.

\section{age and metallicity distribution}\label{app:age_metal_dist}
\setcounter{figure}{0} 

\begin{figure*}
    \centering
    \includegraphics[width=1.7\columnwidth]{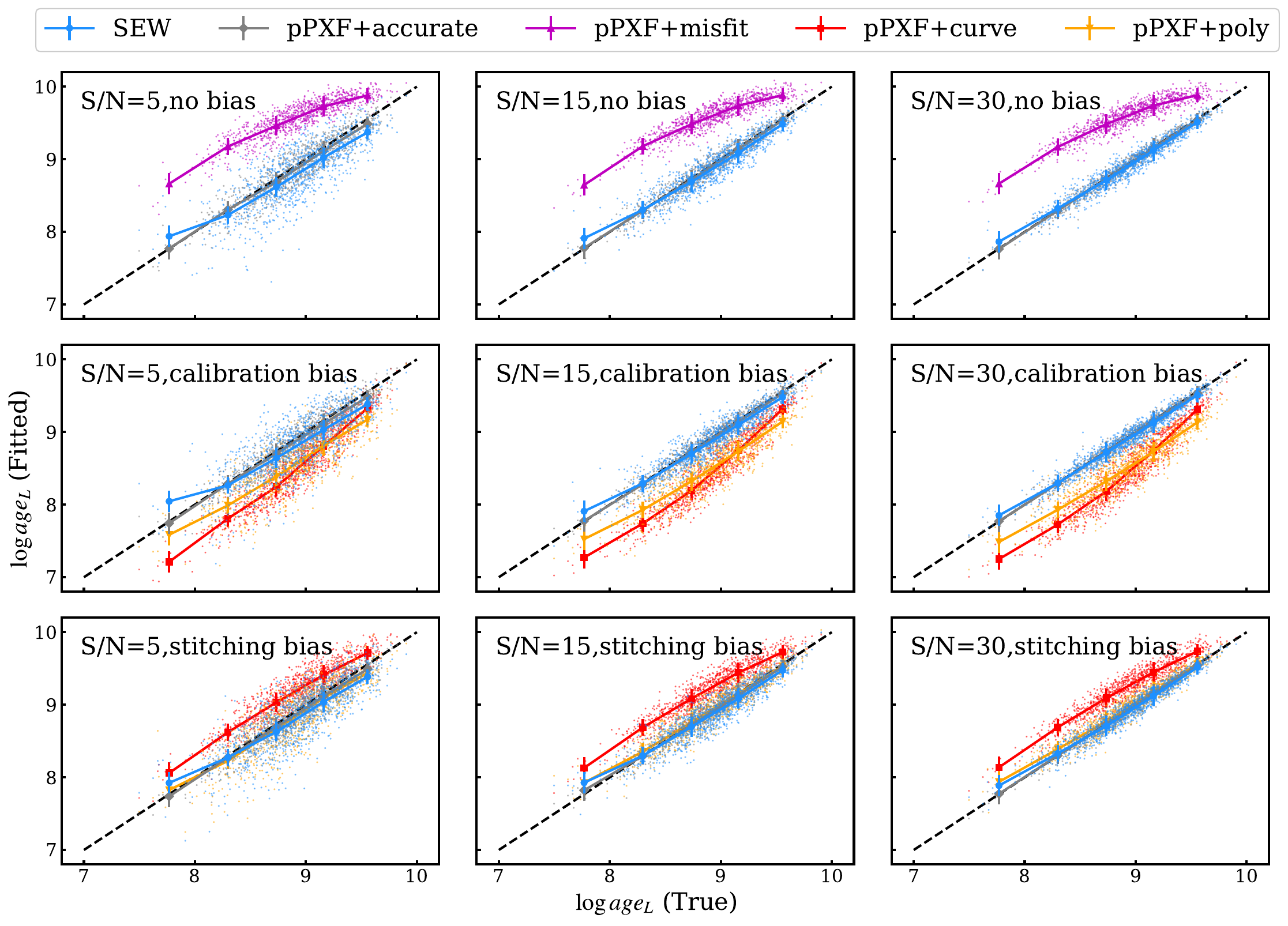}   
    \includegraphics[width=1.7\columnwidth]{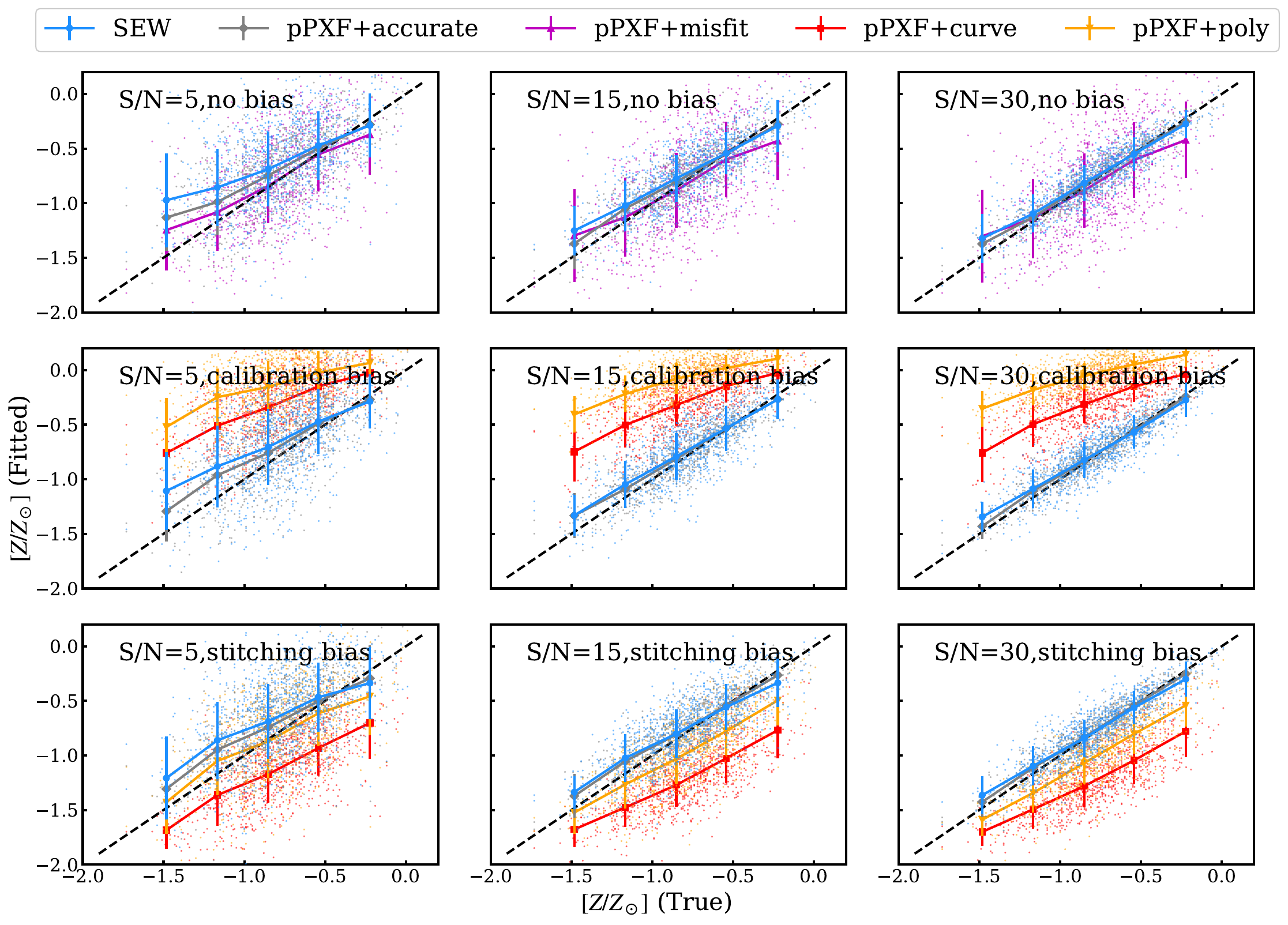}
    \caption{The age (top) and metallicity (bottom) comparison between fitted and True means for different methods and various S/N and flux bias settings. Each subplot displays a specific test condition, labeled in the upper left corner of the plot. Different methodologies are colour-coded in the panel legend, with matching colored error bars indicating binned average measurements along the parameter space.}
    \label{fig:mock_age_metal}
\end{figure*}

This appendix presents supplementary material for the tests described in Section \ref{sec:stat}. Top and bottom panels of figure \ref{fig:mock_age_metal} respectively demonstrate comparisons between the recovered stellar population parameters (age and metallicity) and their true values under different test scenarios, employing various methods according to Section \ref{sec:stat}. 

\bibliographystyle{mnras}
\bibliography{ref} 



\bsp	
\label{lastpage}
\end{CJK*}
\end{document}